\documentclass[a4paper,11pt]{article}
\pdfoutput=1 

\usepackage{jheppub} 

\usepackage[T1]{fontenc} 
\usepackage[utf8x]{inputenc}
\usepackage{color}
\usepackage{stmaryrd, mathrsfs, dsfont, mathtools, hyperref}

\newcommand{\BV}[2]{\left(#1,#2\right)}
\newcommand{\Id}{1\!\!1}

\title{\boldmath 
Uniqueness of ${\cal N}=2$ and $3$ pure supergravities in 4D}
\author[a,1]{Nicolas Boulanger,}\note{Senior Research Associate  
of the F.R.S.-FNRS (Belgium)}
\author[b]{Bernard Julia}
\author[a,2]{and Lucas Traina}\note{Research Fellow of the F.R.S.-FNRS (Belgium)}

\affiliation[a]{Group of Mechanics and Gravitation, Physique th\'eorique et math\'ematique\\
University of Mons -- UMONS, 20 Place du Parc, 7000 Mons, Belgium}
\affiliation[b]{LPT-ENS, PSL and CNRS, 24 rue Lhomond, 75231 Paris CEDEX, France}

\emailAdd{nicolas.boulanger@umons.ac.be, bjulia@lpt.ens.fr, lucas.traina@umons.ac.be}

\abstract{
After proving the impossibility of consistent non-minimal coupling 
of a real Rarita-Schwinger gauge field to electromagnetism, 
we re-derive the necessity of introducing the graviton in order to couple a complex
Rarita-Schwinger gauge field to electromagnetism, with or without a cosmological 
term, thereby obtaining ${\cal N}=2$ pure supergravity 
as the only possibility. 
These results are obtained with the BRST-BV deformation method around 
the flat and (A)dS backgrounds in 4 dimensions.
The same method applied to $n_{v}$ vectors, ${\cal N}$ real spin-3/2 gauge fields
and at most one real spinor field also requires
gravity and yields ${\cal N}=3$ pure supergravity as well as 
${\cal N}=1$ pure supergravity coupled to a vector supermultiplet, 
with or without cosmological terms.
Independently of the matter content, 
we finally derive strong necessary quadratic constraints 
on the possible gaugings for an \textit{arbitrary} number of 
spin-1 and spin-3/2  gauge fields, that are relevant for larger supergravities.
}

\begin{document} 
\maketitle
\flushbottom

\section{Introduction}
\label{sec:intro}

The problem of consistent couplings between a complex 
massless spin-$\tfrac{3}{2}$ field and 
electromagnetism in 4D has been studied long ago 
\cite{Velo:1969bt, Johnson:1960vt}. 
It was re-investigated very recently \cite{Adler:2015yha,Adler:2015zha} 
by Adler who argued that the gauge invariance of the massless 
Rarita-Schwinger\footnote{In 
this paper, by ``massless Rarita-Schwinger field'' we actually mean a 
spin-$\tfrac{3}{2}$ gauge field in the Majorana representation. 
In the original paper by Rarita and Schwinger \cite{Rarita:1941mf}, 
the spinor, whether massive or massless, is in the Dirac 
representation. Throughout this paper, a spinor in the Majorana 
representation will be called \emph{real}, although we do not specify 
the representation used for the Dirac matrices.} 
field could harmlessly be lost in the minimal coupling to
electromagnetism.
However, in \cite{Adler:2017lki} 
it was rigorously shown that for the model \cite{Adler:2015yha,Adler:2015zha}
gauge invariance is crucial. 
More generally, it appears that gauge invariance can 
only be deformed and should never be reduced or augmented 
in the process of introducing perturbative interactions 
among a given set of gauge fields. 
Keeping gauge invariance of constant size, which one calls 
\textit{consistency}, is our main assumption, together with locality, 
no more than two derivatives in the Lagrangian  
and perturbative deformations.
Slightly more precisely, we require that the number of 
gauge parameters is preserved, although the form of the gauge 
transformation laws can be deformed perturbatively.
\vspace{.3cm}

In this paper we start a systematic investigation, 
in four spacetime dimensions, of all the possible consistent couplings 
of a set of $\mathcal{N}_4\geqslant 1$ massless spin-$3/2$ fields and 
a set of $n_v$ vector gauge fields.
We also allow, if necessary, for the introduction of at most one graviton and 
at most one real spin-$\tfrac{1}{2}$ field. 
On the technical side, we follow the cohomological approach 
of \cite{Barnich:1993vg} based on the antifield \cite{Batalin:1981jr,Batalin:1984jr} 
reformulation of the perturbative deformation procedure exposed 
in \cite{Berends:1984rq}.
We refer to \cite  {Boulanger:2000rq} where this framework was applied to deform 
a sum of Fierz--Pauli actions and a pedagogical exposition was given 
on the general procedure. See e.g.  
\cite{Henneaux:2012wg,Barnich:2017nty,Bizdadea:2015yip,Henneaux:2013gba} for recent works 
where this approach has been followed. 
At first order in the infinitesimal deformation parameters 
one typically finds linear constraints restricting the number of possible 
couplings. At second order in the parameters one obtains quadratic constraints 
among them.
\vspace{.3cm}

In this note, among other things we re-derive and strengthen the old result 
on the impossibility of consistent coupling (minimal or non minimal)   
of one real Rarita-Schwinger field to electromagnetism. 
Then, starting with two real Rarita-Schwinger gauge fields coupled to 
Maxwell fields, we see that 
the introduction of the graviton is necessary in order 
to ensure consistency at second order in the infinitesimal deformation parameters.
Starting from three Rarita-Schwinger gauge fields coupled 
to vector gauge fields, this time it is 
the introduction of both the graviton and one Majorana spin-$\tfrac{1}{2}$ field
that is necessary in order to ensure the consistency of the coupling 
at second order.
As a matter of fact, in these two situations
the coupling constant $\kappa$ gets related to  Newton's constant
$G_N=\frac{\kappa^2}{32\pi }\,$, where we have set $c=1\,$.
Thereby, the consistency of the couplings of massless spin-$\tfrac{3}{2}$ fields 
to vector gauge fields naturally leads to the uniqueness of ${\cal N}_4 = 2$
and ${\cal N}_4 = 3$ pure supergravities in 4 dimensions, 
with and without cosmological constant terms. 
Restricting the matter spectrum to contain at most one spin-$\tfrac{1}{2}$ 
field and no scalars excludes the theories of supergravity with 
${\cal N}_4\geqslant 4$ and excludes the 
matter-coupled supergravities, except for ${\cal N}_4=1$ supergravity 
coupled to one vector $(1,1/2)$ multiplet~\cite{Ferrara:1976um}.
\vspace{.3cm}

Still, independently of the restriction of the
scalar and spin-$\tfrac{1}{2}$ field contents, 
we derive strong necessary constraints 
on the possible gaugings for an \textit{arbitrary} number of 
massless spin-$\tfrac{3}{2}$ and spin-1 fields.
We defer the analysis of couplings with more than one Majorana 
spin-$\tfrac{1}{2}$ field and some scalar fields to another paper. 
It is anticipated that the introduction of several Majorana spinor fields 
together with some scalar fields will be necessary 
to ensure consistency of pure $\mathcal{N}_4 \geqslant 4$
supergravity and of matter-coupled supergravities. 
An appealing long term goal would be to automatize 
the computation of the relevant cohomologies 
for the construction and classification of solutions.
\vspace{.3cm}

We exploit the exhaustivity of the cohomological 
reformulation \cite{Barnich:1993vg} of the consistent 
coupling problem in the antifield formalism of Batalin and Vilkovisky 
\cite{Batalin:1981jr,Batalin:1984jr} (BV, for short) in order to 
classify all the possible couplings involving 
a set of $\mathcal{N}_4\,$ spin-$\tfrac{3}{2}$ gauge fields, 
$n_v$ Maxwell vectors, a single massless spin-2 field --- if necessary --- 
together with matter in the form of at most one  Majorana spinor.
Some of our results can be summarised in the form of a theorem and a corollary. 
\vspace{.5cm}

\textbf{Theorem.} [Uniqueness of ${\cal N}_4=2$ and ${\cal N}_4=3$ supergravities]
\textit{With the assumptions of locality, Poincar\'e invariance, 
parity invariance, no more than two derivatives in the deformed Lagrangian and 
at most one Majorana spin-$\tfrac{1}{2}$ ``matter'' field, 
the ${\cal N}_4=2$ and ${\cal N}_4=3$ supergravity theories of 
\cite{Freedman:1976aw,Fradkin:1976xz} 
are the unique ways to consistently couple at least two   
real spin-$\tfrac{3}{2}$ gauge fields to vector gauge fields around 
4D Minkowski spacetime. 
In particular, the cosmological constant terms appear as consistent 
deformations.
Neither local supersymmetry, 
nor general covariance nor even minimal coupling are assumed:
They all appear as a result of the consistency of the deformation.
In particular, dynamical gravity and its diffeomorphism invariance are 
necessarily required, thereby leading to background independence.} 
\vspace{.5cm}

The zero cosmological constant limit of the ${\cal N}_4=2$ 
and ${\cal N}_4=3$ $AdS_4$ supergravity theories of 
\cite{Freedman:1976aw,Fradkin:1976xz} is smooth and 
gives rise to the two earlier models \cite{Ferrara:1976fu} and \cite{Freedman:1976nf}.
A priori, there could have been other theories coupling two and three 
Rarita-Schwinger gauge fields to electromagnetism and to $SO(3)$ 
Yang--Mills gauge fields. In fact, there are no others and this follows 
from our theorem. A corollary can be stated as follows:
\vspace{.3cm}

\textbf{Corollary.}
\textit{The only consistent way to couple
the ${\cal N}_4=1\,$ massless multiplet 
$\left( 1, \frac{3}{2} \right)$   to the 
${\cal N}_4=1\,$ gravitational multiplet
$\left( \frac{3}{2} , 2 \right)$ is through ${\cal N}_4=2\,$ 
supergravity. Analogously, the only consistent way to couple
the ${\cal N}_4=2\,$ massless multiplet 
$\left(\frac{1}{2}, 1, 1, \frac{3}{2} \right)$  to the 
${\cal N}_4=2\,$ gravitational multiplet
$\left( 1, \frac{3}{2} ,\frac{3}{2} , 2 \right)$ 
is through ${\cal N}_4=3\,$ supergravity.}
\vspace{.3cm}

In particular, our results reproduce the well-known 
impossibility to minimally couple some massless 
spin-$\tfrac{3}{2}$ fields to 
electromagnetism around flat Minkowski spacetime.
As was found long ago in
\cite{Freedman:1976aw,Fradkin:1976xz}, the solution 
to this problem requires the introduction of a (negative) 
cosmological constant as well as dynamical gravity. 
As stated in our theorem, the theories with a nonvanishing 
cosmological constant are obtained from 
theories without any cosmological constant,  
starting from flat spacetime, 
where infinitesimal cosmological constant terms 
appear as deformations that can be continued to all orders.
As was shown in \cite{Boulanger:2000rq} in the case of 
massless spin-2 fields around Minkowski spacetime, 
an infinitesimal cosmological constant term appears as a 
consistent, first-order deformation
that can be continued to all orders, without 
introducing any explicit Cartesian coordinate $x^\mu$ 
dependence at any stage of the deformation process, 
leading to the Einstein-Hilbert Lagrangian with finite 
cosmological constant term. 
In this sense, our theorem stated above is the natural 
generalisation of the theorems proved in 
\cite{Boulanger:2000rq} and in \cite{Boulanger:2001wq}.
\vspace{.3cm}

In the present paper, however, our strategy is slightly different, 
although it leads to the same end result stated in our above theorem: 
In the presence of massless spin-$\tfrac{3}{2}$ fields and 
waiving the requirement of Poincar\'e invariance 
(in particular, allowing for explicit $x^\mu$ dependence in
the deformed Lagrangian), the infinitesimal cosmological-constant 
deformation of the quadratic 
Lagrangian around flat spacetime can be continued to all 
orders to give the \textit{quadratic} action consisting of the 
sum of the Fierz--Pauli action in $AdS_4$ with cosmological 
constant term and several Rarita--Schwinger
actions with mass-like terms around $AdS_4\,$.
That the cosmological constant must be negative
was re-derived in \cite{Boulanger:2001wq}, consistently 
with the original finding of \cite{Townsend:1977qa}.
Then, from this quadratic Lagrangian in $AdS_4$ background, 
we pursue the process of consistent deformation, this time looking 
for all the possible $SO(2,3)\,$-invariant cubic vertices 
around $AdS_4\,$.
The advantage of this strategy is that the cosmological constant 
terms are finite from the very beginning and the minimal coupling terms 
arise at the first order in deformation, instead of appearing at second
order as it is the case when one sticks to Minkowski background and 
Poincar\'e invariance at every stage.
The end result is the same in both ways: Gravity is required
for consistency of the deformation, and with it, diffeomorphism 
invariance that washes away the relevance of the initial background.
\vspace{.3cm}

We actually obtained other results that are not contained in our theorem:
(A) Whenever the Majorana spin-$\tfrac{1}{2}$ field is required for 
the coupling of massless spin-$\tfrac{3}{2}$ fields to vector gauge
fields, the spin-$\tfrac{1}{2}$ field must be massless for consistency.  
(B) If one relaxes the assumption that there should be at least two 
real spin-$\tfrac{3}{2}$ gauge fields in the spectrum, we found 
that altogether three interacting models can arise for the coupling of 
the Maxwell fields to the spin-$\tfrac{3}{2}$ gauge fields:

\begin{itemize}
    \item[(B.1)] One of the $n_v$ vectors is coupled to one spin-$\tfrac{3}{2}$ gauge fields, 
the spin-$\tfrac{1}{2}$ field and gravity according to the theory 
\cite{Ferrara:1976um}. This model describes the coupling of a 
(gauged) ${\cal N}_4=1$ supermultiplet $(2,3/2)$ to a (rigid) ${\cal N}_4=1$ 
vector supermultiplet $(1,1/2)\,$. 
The other spin-$\tfrac{3}{2}$ gauge fields remain decoupled from all the rest.
We found that the model in \cite{Ferrara:1976um} can be extended 
so as to include the cosmological constant terms, as in \cite{Townsend:1977qa} 
or in the $SO(3)$ model 
of \cite{Freedman:1976aw};\footnote{We could not 
find any reference mentioning this extension of the model \cite{Ferrara:1976um}, 
although it must be known to experts.}

\item[(B.2)] Only one of the $n_v$ vectors, 
that is identified with the factor $U(1)$ of electromagnetism, 
is coupled to two of the spin-$\tfrac{3}{2}$ gauge fields
through pure ${\cal N}_4 = 2$ supergravity, with and without
cosmological constant terms \cite{Ferrara:1976fu,Freedman:1976aw}, 
the other $n_v-1\,$ vectors remaining decoupled from the 
remaining ${\cal N}_4-2\,$ spin-$\tfrac{3}{2}$ gauge fields; 

\item[(B.3)] Only three of the $n_v$ vectors, that will be identified with an $SO(3)\,$-valued Yang-Mills connection, 
are coupled to three of the Rarita-Schwinger gauge fields, 
minimally and non-minimally, with and without cosmological constant terms 
\cite{Freedman:1976nf,Freedman:1976aw}.

\end{itemize}

Finally, (C) Strong necessary constraints, valid regardless of the 
number of vectors and Rarita-Schwinger gauge fields, 
are found on the possible gaugings; they are independent of the matter 
content of the spectrum.
\vspace{.3cm}

The plan of the paper is as follows. 
In the next section \ref{sec:CohomoRefor}, 
we very briefly review the cohomological 
framework \cite{Barnich:1993vg} for introducing 
consistent interactions in a gauge theory. 
In section \ref{sec:Consistent}, we apply these techniques and the 
methodology of \cite{Barnich:1994db,Barnich:1994mt} 
in order to investigate the possible couplings between 
a set of real Rarita-Schwinger gauge fields and several 
vector gauge fields, possibly with gravity and a real 
spin-$\tfrac{1}{2}$ field. 
Section \ref{sec:Consistent} leads, as main result, 
to the uniqueness of ${\cal N}_4=2$ and ${\cal N}_4=3$ 
pure supergravities, with and without 
cosmological constant terms. 
In subsection \ref{subsection:3.4}
we also derive general quadratic constraints on possible gaugings.
These are necessary constraints that must be satisfied independently 
of the matter content of the undeformed Lagrangian and are therefore 
also valid for extended supergravities with ${\cal N}_4\geqslant 4\,$. 
Section \ref{sec:Outlooks} gives some outlooks and conclusions. 
In a first appendix we gather some identities on spinors 
and Clifford algebra, while the second appendix comments the 
results of the paper in terms of the original Noether 
procedure for introducing interactions from gaugings of 
rigid (non-abelian) symmetries. 
The global symmetries being ``gauged'' disappear as such in the 
deformation: They are subsumed in the emerging non-abelian gauge group.
The BRST-BV approach followed in the body of the text is different, 
as it starts from abelian \emph{gauge} symmetries that are then 
consistently deformed to result in non-abelian gauge theories. 

\section{Cohomological reformulation of the deformation problem}
\label{sec:CohomoRefor}

In this section, we briefly review the cohomological procedure 
\cite{Barnich:1993vg} for perturbative deformation of a Lagrangian 
gauge theory, exemplifying it on the free theories describing 
massless spin-$s$ fields around flat and $AdS_4$ backgrounds for 
$s\in \{\frac{1}{2}, 1,\frac{3}{2}, 2\}\,$.
We also present our conventions and notation for spinors. 

\subsection{Cohomological approach}

\paragraph{Initial theory.}
In order to introduce consistent interactions to an initial theory characterised by 
an action $S_0[\varphi^i]$ invariant under the gauge transformations\footnote{We use 
De Witt's condensed notation.} 
$\Delta_0 \varphi^i = {R_0}^i{}_{\alpha}\,\epsilon^\alpha\,$, an appropriate
framework is provided by the reformulation \cite{Barnich:1993vg} of 
the Noether deformation procedure in the Batalin-Vilkovisky (BV) antifield formalism
\cite{Batalin:1981jr,Batalin:1984jr}.
An advantage of the antifield formalism is that the whole gauge 
structure, including initial action $S_0[\varphi^i]=\int d^4x \,{\cal L}_0\,$, 
its gauge invariances and the gauge algebra, is captured by the 
BV master action (or BV functional) $W_0\,$, see 
e.g.~\cite{Henneaux:1992ig,Gomis:1994he} for reviews. 
The master action is a functional 
\begin{equation}
  \label{eq:masteraction}
  W_0[\varphi^i,C^\alpha,\varphi^\star_i,C^\star_\alpha] 
  = S_0[\varphi^i] + \varphi^\star_i {R_0}^{i}{}_\alpha \, 
  C^\alpha + \tfrac{1}{2}\,C^\star_\alpha
  {f_0}^\alpha{}_{\beta\gamma} C^\beta C^\gamma  + \dots \;, 
\end{equation}
that satisfies the classical \textit{master equation}
\begin{equation}
\label{eq:masterequation}
 (W_0,W_0)=0\;.
\end{equation}
In this equation, the BV antibracket is the graded-odd Lie bracket 
defined by
\begin{equation}
  \label{eq:antibracket}
  (X,Y) \coloneqq \int\!d^4\!x\, \left[ \frac{\delta^R X}{\delta \Phi^A(x)}
\frac{\delta^L Y}{\delta \Phi^\star_A(x)}-\frac{\delta^R X}{\delta 
\Phi^\star_A(x)}\frac{\delta^L Y}{\delta \Phi^A(x)} \right]\;,
\end{equation}
on the extended space locally coordinatised by 
$\Phi^A\coloneqq(\varphi^i,C^\alpha)\,$, the original
fields and ghosts for the type of theories we will be interested in,  
and their canonically conjugated antifields $\Phi^\star_A=(\varphi^\star_i,C^\star_\alpha)\,$. 
The Lagrangian, gauge 
variations of the fields and structure functions of the gauge 
algebra are contained in the first, second and third term of the 
master action \eqref{eq:masteraction}, respectively. 
The BRST differential $s\,$ associated with the 
initial theory, acts as
\begin{align}
    s\;\bullet  \coloneqq (W_0,\bullet)\;.
\end{align}
That it is a differential follows from the master 
equation \eqref{eq:masterequation}. 
A $\mathbb{Z}$ grading is associated with $s\,$, that is called the ghost number 
\emph{gh}. The BRST differential increases the ghost number by one unit. 
For the type of theories we are interested in, $s$ decomposes into the sum of 
two differentials: $s=\delta + \gamma\,$. 
For a free theory action $S_0\,$, the Koszul-Tate 
differential $\delta$ acts only on the antifields $\Phi^\star_A\,$, 
while $\gamma\,$, the differential along the gauge orbits,  
acts only on the fields $\Phi^A\,$. 
Since the initial theory is free, the structure functions 
${f_0}^\alpha{}_{\beta\gamma}$ vanish, 
therefore $\gamma$ only acts on the fields $\varphi^i$ of 
the quadratic action $S_0\,$. Together with $\delta^2= 0 = \gamma^2\,$, 
the anticommutation relation $\{\delta,\gamma\}=0$ is true. 
One extends the 
action of $s$ on derivatives of the fields and antifields by $\{s ,d\}=0\,$, 
where $d$ is the total exterior derivative 
\begin{align}
d \coloneqq dx^\mu \partial_\mu = dx^\mu 
\left( \frac{\partial}{\partial x^\mu}\, 
+ \Phi^A{}_\mu \,\frac{\partial^L}{\partial \Phi^A} 
+ {\Phi^{\star}_A}_\mu\, \frac{\partial^L}{\partial \Phi^\star_A} + \ldots \right)\;,
\end{align}
where $\Phi^A{}_\mu = \frac{\partial \Phi^A}{\partial x^\mu}\,$, idem for 
${\Phi^\star_A}_\mu\,$. The Koszul-Tate differential decreases the antifield number 
\textit{antifd} by one unit, while the differential $\gamma$ increases the pure ghost
number \textit{puregh} by one unit, so that one has $gh = puregh - antifd\,$. 
The BV functional $W_0$ is required to have a
definite ghost number and to 
start with the initial action,  $W_0 = S_0 ~+ $ antifield-dependent terms, 
therefore $gh(W_0)=0\,$ since one assigns zero ghost number 
to the fields $\varphi^i\,$.
\vspace{.3cm}

In the BRST formalism, the ghosts $C^\alpha$ have the opposite Grassmann parity 
compared to the original gauge parameters $\epsilon^\alpha\,$. 
In the antifield extension of the BRST formalism, 
the Grassmann parities of the antifields are given by 
$|\Phi^\star_A| = |\Phi^A| + 1\,$, where $|Z| = 0$ mod $2$ for 
bosonic fields $Z$, and $|Z| = 1$ mod $2$ for fermionic fields $Z\,$. 
Moreover, the ghost number of the antifields is
$gh(\Phi^\star_A) = -gh(\Phi^A)-1\,$. 
One demands that $W_0$ be of definite Grassmann
parity, therefore $|W_0| = 0\,$.

\paragraph{Infinitesimal deformations.}

The antifield formalism is also particularly efficient  
in order to perturbatively deform the initial action \cite{Barnich:1993vg}, 
$S[\varphi^i] = S_0 + S_1+S_2+\ldots \,$, where $S_1$ and $S_2$ are 
respectively linear and quadratic in the collection  
of infinitesimal parameters, 
while at the same time deforming the initial gauge variations
$\Delta = \Delta_0 + \Delta_1+\Delta_2+\ldots\,$, 
such that one has $\Delta S = 0\,$.  
We introduce a set of infinitesimal deformation parameters collectively 
denoted by $g$ and relative coupling constants $\tilde\kappa$ 
(index omitted for both $g$ and $\tilde\kappa$) 
so that the expansion of $S$ is in the $g\tilde{\kappa}$’s.
The requirement of gauge invariance $\Delta S = 0$
amounts to asking that 
$W[\Phi^A,\Phi^\star_A] = W_0[\Phi^A,\Phi^\star_A] + g\,W_1[\Phi^A,\Phi^\star_A] + g^2\,W_2[\Phi^A,\Phi^\star_A] + {\cal O}(g^3)$ 
satisfies the master equation \eqref{eq:masterequation}. 
To first and second orders in the deformation parameters, it means that
\begin{align}
    (W_0, W_1) &= 0\;,\label{eq:firstorderdef}\\
    (W_0, W_2) + \tfrac{1}{2}\,(W_1, W_1) &= 0 \label{eq:secondorderdef} \;. 
\end{align}
Upon recognising that an infinitesimal deformation $W_1$ is trivial
if it is of the form $W_1 = (W_0,B) = sB$ for a local functional $B$ of ghost number 
$-1\,$, the classification of infinitesimal deformations is a cohomological 
problem, see \cite{Barnich:1993vg} and \cite{Boulanger:2000rq} for more details. \vspace{.3cm}

We assume that the fields and their derivatives vanish at infinity, 
so that we do not take boundary terms into account inside the action 
or its variation. 
Decomposing the first order infinitesimal deformation $W_1$ with respect to the antifield 
number, general results derived in \cite{Barnich:1994db,Barnich:1994mt} 
based on the assumption of locality of the deformation 
allow us to stop at antifield number 2 for the type of theories at hand:
\begin{align}
    W_1 = \int d^4x\, (a_0 + a_1 + a_2 )\;.
\end{align}
Equation \eqref{eq:firstorderdef} then yields 
a $(\delta, \gamma)$ descent of equations in  antifield numbers 
for the densities $a_{antifd}\,$ (from $a_2\,$ to $a_1\,$ and $a_0\,$):
\begin{align}
\label{da1ga0}
\delta a_1 + \gamma a_0 &= \partial_\mu j^\mu_0\;,\\
\label{da2ga1}
\delta a_2 + \gamma a_1 &= \partial_\mu j^\mu_1\;, \\
\label{ga20}
\gamma a_2 &= 0\;.
\end{align}
That one can redefine away a total derivative from the right-hand side
of the last equation 
also follows from standard arguments \cite{Barnich:1994db,Barnich:1994mt}.

At second order in deformation, for 
$W_2 = \int d^4x\, (b_0 + b_1 + b_2)\,$, 
the equation \eqref{eq:secondorderdef} splits into
\begin{align}
\label{db1ga0a1a0}
\delta b_1 + \gamma b_0 &= - ({a_1},{a_0}) + \partial_\mu t^\mu_0\;,\\
\label{sgammaa1}
\delta b_2 + \gamma b_1 &= -\tfrac{1}{2}({a_1},{a_1}) - ({a_2},{a_1}) + \partial_\mu t^\mu_1\;,\\
\label{a2a2}
\gamma b_2 &= - \tfrac{1}{2}\,({a_2},{a_2}) + \partial_\mu t^\mu_2\;.
\end{align}
The form of the right hand sides of these second order equations is restricted by properties of the solutions of the first order ones.

\subsection{First starting point: Free BV functionals in flat background}

The BV spectrum of the theories generated by $\mathcal{N}_4$ 
massless Rarita-Schwinger fields and $n_v$ vector gauge fields is given by :
\begin{itemize}
\item[$\bullet$] 1 graviton whose gauge potential is $h_{\mu\nu}\,$, 
the metric perturbation around the Minkowski background, such that  
$g_{\mu\nu} = \eta_{\mu\nu} + \kappa h_{\mu\nu}\,$ is the complete metric. 
(In the $AdS_4$ background, $h_{\mu\nu}$ will refer to the 
decomposition $g_{\mu\nu} = \bar{g}_{\mu\nu} + \kappa\, h_{\mu\nu}$ 
of the dynamical metric around the $AdS_4$ background $\bar{g}_{\mu\nu}\,$.);

\item[$\bullet$] $\mathcal{N}_4$ gravitini $(\psi^\Delta_{\mu})_A\,$, $\Delta=1,\dots,\mathcal{N}_4\,$ in the Majorana representation, 
where we refer to Appendix \ref{Diracology} for the conventions 
used for 4-component Majorana spinors with index $A\,$;
\item[$\bullet$] $n_v$ Maxwell gauge fields $A_\mu^a\,$, $a=1,\dots,n_v\,$;
\item[$\bullet$] Zero or one spin-$\frac{1}{2}$ field $(\chi)_A$ in the Majorana representation;
\item[$\bullet$] The ghosts $\xi_\mu$ associated with linearised diffeomorphisms;
\item[$\bullet$] $\mathcal{N}_4$ ghosts $\zeta^\Delta$ associated with the spin-$\tfrac{3}{2}$ 
gauge symmetry of each Rarita-Schwinger gauge fields;
\item[$\bullet$] $n_v$ ghost(s) $\mathcal{C}^a$ associated with 
the abelian factors $(U(1)){}^{\times n_v}$ of the gauge group;
\item[$\bullet$] An antifield $\Phi^\star_I$ associated with each field or ghost $\Phi^I\,$.
\end{itemize}
Each of these fields carries a pure ghost number \emph{puregh}, an 
antifield number \emph{antifd} and a ghost number 
$gh=puregh - antifd\,$ 
that are listed in table \ref{tableFlat} and in table \ref{tableAdS}, 
in which we also indicated the action of the differentials 
$\gamma$, $\delta$ and $s=\gamma+\delta\,$ around the flat and 
$AdS_4$ backgrounds. 

In Minkowski background, the BV functional of the free theory associated with 
the sum of the free actions for a massless spin-2, several spin-$\tfrac{3}{2}$ and spin-1 
fields and one spin-$\tfrac{1}{2}$ field is given by
\begin{align}
{W}_0 &= S^D[\chi_A] + S^M[A_\mu^a] + S^{RS}[\psi^\Delta_{\mu A}] + S^{FP}[h_{\mu\nu}] \notag\\
&~~~ +\int d^4x\, A^{\star\mu}_{\phantom{\star}a} \partial_\mu \mathcal{C}^a + \int d^4x\, \psi^{\star\mu A}_{\phantom{\star}\Delta} \partial_\mu \zeta^\Delta_A + \int d^4x\, 2 h^{\star\mu\nu} \partial_{(\mu} \xi_{\nu)}\;,
\label{masterequflatspace}
\end{align}
where
\begin{align}
S^D[\chi_A] &= -\tfrac{1}{2} \int d^4x\, \bar{\chi}^A (\gamma^\mu)_A^{\phantom{A}B} \partial_\mu \chi_B\;,\\
S^M[A_\mu^a] &= -\tfrac{1}{4} \int d^4x\, F_{\mu\nu}^a F^{b\mu\nu} \delta_{ab},~~~~~F_{\mu\nu}^a \coloneqq \partial_\mu A_\nu^a - \partial_\nu A_\mu^a\;,\\
S^{RS}[\psi^\Delta_{\mu A}] &= -\tfrac{1}{2} \int d^4x\, \bar{\psi}^{\Delta A}_\mu (\gamma^{\mu\nu\rho})_A^{\phantom{A}B} \partial_\nu \psi^\Omega_{\rho B}  \delta_{\Delta\Omega}\;,\\
S^{FP}[h_{\mu\nu}] &= \int d^4x\, \left( -\tfrac{1}{2} \partial_\mu h_{\nu\rho} \partial^\mu h^{\nu\rho} + \partial_\mu h^\mu_{\phantom{\mu}\nu} \partial_\rho h^{\rho\nu} - \partial_\nu h_\mu^{\phantom{\mu}\mu} \partial_\rho h^{\rho\nu} + \tfrac{1}{2} \partial_\mu h_\nu^{\phantom{\nu}\nu} \partial^\mu h_\rho^{\phantom{\rho}\rho} \right)\;.
\end{align}
The field equations for the Fierz-Pauli gauge field $h_{\mu\nu}$ 
are denoted by $-2\overset{(1)}{G}{}^{\mu\nu}\approx 0\,$, where 
weak equality means ``on the surface of the solutions to the free field equations'' 
and where the linearised Einstein tensor
is given by 
\begin{align}
\overset{(1)}{G}{}^{\mu\nu} \coloneqq -\tfrac{1}{2}\left(
\Box h^{\mu\nu} - \partial^\mu \partial_\rho h^{\rho\nu} - \partial^\nu \partial_\rho h^{\rho\mu} + \partial^\mu \partial^\nu h + \eta^{\mu\nu} \partial_\rho \partial_\sigma h^{\rho\sigma} - \eta^{\mu\nu} \Box h \right)\;.
\end{align}
The action of the BRST differential acting on each field and antifield 
is summarised in Table \ref{tableFlat}.
\begin{center}
\begin{table}
\begin{tabular}{|c||c|c|c|c||c|c|c|c|}
\hline
 & $|.|$ & $gh$ & $puregh$ & $antifd$ & $\gamma$ & $\delta$ & $s=\gamma+\delta$\\
 \hline
 \hline
 $\chi_A$ & 1 & 0 & 0 & 0 & 0 & 0 & 0\\
 \hline
$A_\mu^a$ & 0 & 0 & 0 & 0 & $\partial_\mu \mathcal{C}^a$ & 0 & $\partial_\mu \mathcal{C}^a$\\
\hline
$\psi^\Delta_{\mu A}$ & 1 & 0 & 0 & 0 & $-\partial_\mu \zeta^\Delta_A$ & 0 & $-\partial_\mu \zeta^\Delta_A$\\
\hline
$h_{\mu\nu}$ & 0 & 0 & 0 & 0 & $2 \partial_{(\mu} \xi_{\nu)}$ & 0 & $2 \partial_{(\mu} \xi_{\nu)}$\\
\hline
\hline
$\mathcal{C}^a$ & 1 & 1 & 1 & 0 & 0 & 0 & 0\\
\hline
$\zeta^\Delta_A$ & 0 & 1 & 1 & 0 & 0 & 0 & 0\\
\hline
$\xi_\mu$ & 1 & 1 & 1 & 0 & 0 & 0 & 0\\
\hline
\hline
$\chi^{\star A}$ & 0 & -1 & 0 & 1 & 0 & $\partial_\mu \bar{\chi}^B (\gamma^\mu)_B^{\phantom{B}A}$ & $\partial_\mu \bar{\chi}^B (\gamma^\mu)_B^{\phantom{B}A}$\\
\hline 
$A^{\star\mu}_{\phantom{\star}a}$ & 1 & -1 & 0 & 1 & 0 & $\partial_\nu F^{\nu\mu}_a$ & $\partial_\nu F^{\nu\mu}_a$\\
\hline
$\psi^{\star\mu A}_{\phantom{\star}\Delta}$ & 0 & -1 & 0 & 1 & 0 & $- \partial_\nu \bar{\psi}_{\Delta\rho}^{B} (\gamma^{\mu\nu\rho})_B^{\phantom{B}A}$ & $- \partial_\nu \bar{\psi}_{\Delta\rho}^{B} (\gamma^{\mu\nu\rho})_B^{\phantom{B}A}$\\
\hline
$h^{\star\mu\nu}$& 1 & -1 & 0 & 1 & 0 & $-2\overset{(1)}{G}{}^{\mu\nu}$ & $-2\overset{(1)}{G}{}^{\mu\nu}$\\
\hline
\hline
$\mathcal{C}^\star_a$ & 0 & -2 & 0 & 2 & 0 & $- \partial_\mu A^{\star\mu}_{\phantom{\star}a}$ & $- \partial_\mu A^{\star\mu}_{\phantom{\star}a}$\\
\hline
$\zeta^{\star A}_{\phantom{\star}\Delta}$ & 1 & -2 & 0 & 2 & 0 & $- \partial_\mu \psi^{\star\mu A}_{\phantom{\star}\Delta}$ & $- \partial_\mu \psi^{\star\mu A}_{\phantom{\star}\Delta}$\\
\hline
$\xi^{\star\mu}$ & 0 & -2 & 0 & 2 & 0 & $-2 \partial_\nu h^{\star\mu\nu}$ & $-2 \partial_\nu h^{\star\mu\nu}$\\
\hline
\end{tabular}
\caption{Grassmann parity, ghost number, pureghost number, antifield number 
and actions of the differentials $\gamma$, $\delta$ and $s$ on 
the various fields of the spectrum, for the theory around flat background.}
\label{tableFlat}
\end{table}
\end{center}
A central cohomological group for the deformation procedure of the initial theory 
is $H(\gamma)\,$, the cohomology of $\gamma$ in the space of local functions. 
Defining the field strength
\begin{align}
K_{\alpha\beta|\mu\nu} \coloneqq -
\partial_{\alpha}\partial_{[\mu}  h_{\nu]\beta} 
+\partial_{\beta}\partial_{[\mu}  h_{\nu]\alpha} \;,
\end{align} 
$H(\gamma)\,$ is given by
\begin{equation}
\label{cohomologyMink}
H(\gamma) \cong \left\{ f \left( [\chi_A] ~,~ [F^a_{\mu\nu}] ~,~ [\partial_{[\mu}\psi^\Delta_{\nu] A}] ~,~ [K_{\alpha\beta|\mu\nu}] ~,~ \mathcal{C}^a ~,~ \zeta^\Delta_A ~,~ \xi_\mu ~,~ \partial_{[\mu} \xi_{\nu]} ~,~ [\Phi^\star_I] \right) \right\}
\end{equation}
where the notation $[\Phi]$ means the field $\Phi$ and all its derivatives up to 
some finite, but arbitrary, order. 
The almost complete elimination of the derivatives of the 
ghosts and the appearance of gauge invariant expressions follow 
from the existence of corresponding contractible pairs for contracting 
homotopies. For vector fields, the cohomology $H(\gamma)$ was computed long ago.
We refer the reader to \cite{Barnich:1994mt} for details and references. 
In the massless spin-2 case, the cohomology of $\gamma$ was computed 
in \cite{Boulanger:2000rq}, while in the presence of a Rarita-Schwinger gauge field, 
it was given in \cite{Boulanger:2001wq}. 

\subsection{Second starting point: Anti de Sitter background}
\label{subsec:AdSBack}


The components of the background $AdS_4$ vierbeins and spin connection 
are denoted by $\bar e_{\mu}{}^a$ and $\bar\omega_\mu{}^{ab}\,$. 
The components of the background $AdS_4$ metric therefore read 
$\bar{g}_{\mu\nu} = \bar e_{\mu}{}^a \bar e_{\nu}{}^b\,\eta_{ab}\,$.
We denote the Lorentz-covariant derivative on $AdS_4$ 
by $\nabla_\mu\,$ and the corresponding one-form differential operator 
$\nabla=dx^\mu \nabla_\mu\,$.
Introducing $\sigma_{ab} = \frac{1}{2}\,\gamma_{ab}\,$, then $i\sigma_{ab}\,$ 
gives a representation of the 
Lorentz algebra on the spinor representation and 
$\nabla \psi = d\psi + \Omega \psi\,$ where 
$\Omega \coloneqq \tfrac{1}{4}\,\bar\omega^{ab} \gamma_{ab}\,$.
On the conjugate spinor $\bar \psi\,$, the Lorentz-covariant derivative 
acts like $\nabla \bar\psi = d\bar \psi - \bar \psi\Omega\,$.
Our conventions
are such that the commutator of Lorentz-covariant derivatives on $AdS_4$, 
when acting on a vector or on a spinor, is given by
\begin{align}
[ \nabla_{\mu}, \nabla_{\nu} ] \xi_{\sigma} = -2\lambda^2 \bar{g}_{\sigma[\mu} 
\xi_{\nu]}\;, \quad\quad [ \nabla_{\mu}, \nabla_{\nu} ] \zeta^\Delta_A = - \lambda^2 (\sigma_{\mu\nu})_{A}^{\phantom{A}B} \zeta^\Delta_B\;,
\end{align}
where $\sigma_{\mu\nu} \coloneqq \bar e_\mu{}^a \bar e_\nu{}^b\,\sigma_{ab}\,$, 
$\lambda$ is the $AdS_4$ inverse radius that, in four spacetime dimension, 
is related to the cosmological constant $\Lambda$ by $\Lambda = -3\lambda^2\,$.

In $AdS_4$ background, the BV functional of the free theory associated with 
the sum of the free actions for a massless spin-2, several spin-$\tfrac{3}{2}$ and spin-1 
fields and one spin-$\tfrac{1}{2}$ field is given by
\begin{align}
{W}^\Lambda_0 &= S_\Lambda^D[\chi_A] + S_\Lambda^M[A_\mu^a] + 
S^{RS}[\psi^\Delta_{\mu A}] + S_\Lambda^{FP}[h_{\mu\nu}] 
+\int d^4x\, \sqrt{-\bar{g}}\, A^{\star\mu}_{\phantom{\star}a} \nabla_\mu \mathcal{C}^a 
\notag\\
\label{BVfunctionalAdS}
&~~~ + \int d^4x\, \sqrt{-\bar{g}}\, \psi^{\star\mu A}_{\phantom{\star}\Delta} \left(\nabla_\mu \zeta^\Delta_A + \frac{\lambda}{2} (\gamma_\mu)_A^{\phantom{A}B} \zeta^\Delta_B \right) + \int d^4x\, \sqrt{-\bar{g}}\, 2 h^{\star\mu\nu} \nabla_{(\mu} \xi_{\nu)}\;,
\end{align}
where
\begin{align}
S_\Lambda^D[\chi_A] &= -\frac{1}{2} \int d^4x\, \sqrt{-\bar{g}}\, \bar{\chi}^A (\gamma^\mu)_A^{\phantom{A}B} \nabla_\mu \chi_B\;, \\
S_\Lambda^M[A_\mu^a] &= -\frac{1}{4} \int d^4x \sqrt{-\bar{g}}\, F_{\mu\nu}^a F^{b\mu\nu} \delta_{ab}\;,\\
S_\Lambda^{RS}[\psi^\Delta_{\mu A}] &= \int d^4x \sqrt{-\bar{g}}\, \left( -\frac{1}{2} \bar{\psi}^{\Delta A}_\mu (\gamma^{\mu\nu\rho})_A^{\phantom{A}B} \nabla_\nu \psi^\Omega_{\rho B} + \frac{\lambda}{2} \bar{\psi}^{\Delta A}_\mu (\gamma^{\mu\nu})_A^{\phantom{A}B} \psi_{\nu B}^\Omega \right) \delta_{\Delta\Omega}\;,\\
S_\Lambda^{FP}[h_{\mu\nu}] &= \int d^4x\, \sqrt{-\bar{g}}\, \left( -\frac{1}{2} \nabla_\mu h_{\nu\rho} \nabla^\mu h^{\nu\rho} + \nabla_\mu h^\mu_{\phantom{\mu}\nu} \nabla_\rho h^{\rho\nu} - \nabla_\nu h_\mu^{\phantom{\mu}\mu} \nabla_\rho h^{\rho\nu} \right. \notag\\
&~~~~~~~~~~~~~~~~~~~~~~~~~~+ \left. \frac{1}{2} \nabla_\mu h_\nu^{\phantom{\nu}\nu} \nabla^\mu h_\rho^{\phantom{\rho}\rho} + \frac{\lambda^2}{2} (2h_{\mu\nu} h^{\mu\nu} + h_\mu^{\phantom{\mu}\mu} h_\nu^{\phantom{\nu}\nu})\right)\;.
\end{align}
The gauge-invariant field strengths for the spin-1, spin-$\tfrac{3}{2}$ and
spin-2 fields are
\begin{align}
F_{\mu\nu}^a &\coloneqq \nabla_\mu A_\nu^a - \nabla_\nu A_\mu^a\;,
\quad\quad \Psi^\Delta_{\mu\nu A} \,\coloneqq\, \nabla_{[\mu} \psi_{\nu]A}^\Delta + \frac{\lambda}{2} (\gamma_{[\mu})_A^{\phantom{A}B} \psi^\Delta_{\nu]B}\;,
\\
\tilde K^{\alpha\beta|\mu\nu} &\coloneqq 
-\tfrac{1}{2}\,\left(
\nabla ^\alpha \nabla^{[\mu} h^{\nu]\beta} - \nabla ^\beta \nabla^{[\mu} h^{\nu]\alpha} + \nabla ^\mu \nabla^{[\alpha} h^{\beta]\nu} - \nabla ^\nu \nabla^{[\alpha} h^{\beta]\mu}
\right) + \lambda^2 \left( \bar{g}^{\alpha[\mu} h^{\nu]\beta} - \bar{g}^{\beta[\mu} h^{\nu]\alpha} \right)\, .
\end{align}

When considering the $AdS_4$ background, 
we will denote by $-2 \tilde{G}^{\mu\nu} $ the left-hand side of 
the Euler-Lagrange equations for the Fierz-Pauli gauge field $h_{\mu\nu}\,$. 
The tensor $\tilde{G}^{\mu\nu}$ is the linearisation around $AdS_4$ 
of $G_{\mu\nu} + \Lambda g_{\mu\nu}\,$. It is given by
\begin{align}
\tilde{G}^{\mu\nu} &= -\tfrac{1}{2} \left( \bar{\Box} h^{\mu\nu} - \nabla^\mu \nabla_\sigma h^{\sigma\nu} - \nabla^\nu \nabla_\sigma h^{\sigma\mu} + \nabla^\mu \nabla^\nu h 
+ 
\bar{g}^{\mu\nu} \nabla_\rho \nabla_\sigma h^{\rho\sigma} - \bar{g}^{\mu\nu} \bar{\Box} h + 2\lambda^2 h^{\mu\nu} + \lambda^2 \bar{g}^{\mu\nu} h \right)\;,
\nonumber
\end{align}
where $\bar{\Box} \coloneqq \bar{g}^{\mu\nu} \nabla_\mu \nabla_\nu$ 
is the Laplace-Beltrami operator in $AdS_4$ 
and $h \coloneqq h_{\mu\nu}\bar{g}^{\mu\nu}\,$.
The action of the BRST differential acting on each field and antifield is summarised 
in Table \ref{tableAdS}. Note that, in this table as well as in the action, 
we have not written any mass term for the real spin-$1/2$ field. 
This is not a loss of generality
but a result that is explained in footnote \ref{foonotemass}.
\begin{center}
\begin{table}
\begin{tabular}{|c||c|c|c|c||c|c|c|}
\hline
 & $|\cdot|$ & $gh$ & $puregh$ & $antifd$ & $\frac{1}{\sqrt{-\bar{g}}}\gamma$ & $\frac{1}{\sqrt{-\bar{g}}}\delta$\\
 \hline
 \hline
 $\chi_A$ & 1 & 0 & 0 & 0 & 0 & 0\\
 \hline
$A_\mu^a$ & 0 & 0 & 0 & 0 & $\nabla_\mu \mathcal{C}^a$ & 0\\
\hline
$\psi^\Delta_{\mu A}$ & 1 & 0 & 0 & 0 & $-\nabla_\mu \zeta^\Delta_A - \frac{\lambda}{2} (\gamma_\mu)_A^{\phantom{A}B} \zeta^\Delta_B$ & 0\\
\hline
$h_{\mu\nu}$ & 0 & 0 & 0 & 0 & $2 \nabla_{(\mu} \xi_{\nu)}$ & 0\\
\hline
\hline
$\mathcal{C}^a$ & 1 & 1 & 1 & 0 & 0 & 0\\
\hline
$\zeta^\Delta_A$ & 0 & 1 & 1 & 0 & 0 & 0\\
\hline
$\xi_\mu$ & 1 & 1 & 1 & 0 & 0 & 0\\
\hline
\hline
$\chi^{\star A}$ & 0 & -1 & 0 & 1 & 0 & $\nabla_\mu \bar{\chi}^B (\gamma^\mu)_B^{\phantom{B}A}$ \\
\hline 
$A^{\star\mu}_{\phantom{\star}a}$ & 1 & -1 & 0 & 1 & 0 & $\nabla_\nu F^{\nu\mu}_a$\\
\hline
$\psi^{\star\mu A}_{\phantom{\star}\Delta}$ & 0 & -1 & 0 & 1 & 0 & $- \nabla_\nu \bar{\psi}_{\Delta\rho}^{B} (\gamma^{\mu\nu\rho})_B^{\phantom{B}A} - \lambda \bar{\psi}_{\Delta\nu}^B (\gamma^{\mu\nu})_B^{\phantom{B}A}$\\
\hline
$h^{\star\mu\nu}$& 1 & -1 & 0 & 1 & 0 & $-2\tilde{G}^{\mu\nu}$\\
\hline
\hline
$\mathcal{C}^\star_a$ & 0 & -2 & 0 & 2 & 0& $- \nabla_\mu A^{\star\mu}_{\phantom{\star}a}$ \\
\hline
$\zeta^{\star A}_{\phantom{\star}\Delta}$ & 1 & -2 & 0 & 2 & 0 & $- \nabla_\mu \psi^{\star\mu A}_{\phantom{\star}\Delta} + \frac{\lambda}{2} \psi^{\star\mu B}_{\phantom{\star}\Delta} (\gamma_\mu)_B^{\phantom{B}A}$\\
\hline
$\xi^{\star\mu}$ & 0 & -2 & 0 & 2 & 0 & $-2 \nabla_\nu h^{\star\mu\nu}$\\
\hline
\end{tabular}
\caption{Grassmann parity, ghost number, pureghost number, antifield number 
and actions of the differentials $\gamma$ and $\delta$ on 
the various fields of the spectrum, for the theory around $AdS_4$ background.}
\label{tableAdS}
\end{table}
\end{center}

Finally, along similar lines to those followed in flat space, we find 
that the cohomology of $\gamma$ is
\begin{equation}
H(\gamma) \cong \left\{ f \left( [\chi_A] ~,~ [F^a_{\mu\nu}] ~,~ [\Psi^\Delta_{\mu\nu A}] ~,~ [\tilde{K}^{\alpha\beta|\mu\nu}] ~,~ \mathcal{C}^a ~,~ \zeta^\Delta_A ~,~ \xi_\mu ~,~ \nabla_{[\mu} \xi_{\nu]} ~,~ [\Phi^\star_I] \right) \right\}\,.
\end{equation}
%

\section{Uniqueness of ${\cal N}_4=2$ and ${\cal N}_4=3$
pure supergravities} \label{sec:Consistent}

In this section we first give, in subsection \ref{subsec:ConsistentFlat}, 
the classification of the most general, consistent 
cubic vertices for the initial theory describing 
the free propagation, around the flat background, 
of one graviton, two massless Rarita-Schwinger fields and one Maxwell gauge field. 
The $\mathcal{N}_4 = 2$ pure supergravity theory 
(without cosmological constant) \cite{Ferrara:1976fu} is recovered as 
the unique theory that deforms the gauge algebra away from the abelian 
algebra of the initial, free theory.
In this model, the gravitini are not charged under the $U(1)$ gauge field of 
electromagnetism. The well-known obstruction \cite{Johnson:1960vt,Velo:1969bt} 
to the minimal coupling of the 
gravitini around Minkowski spacetime is rederived in cohomological terms in 
subsection \ref{subsec:Obstru}. 
Most of the technical details of the computations concerning 
subsection \ref{subsec:ConsistentFlat} are not given. 
They can be found in \cite{LucasMasterThesis}, 
and some of them will be given 
later in the next subsection \ref{sec:ConsistentAdS}, where we first 
deform the theory by changing background from flat spacetime to $AdS_4\,$, and 
then further deform the theory until the actions of pure 
$\mathcal{N}_4 = 2$ and $\mathcal{N}_4 =3$ 
supergravities with cosmological constant terms \cite{Freedman:1976aw} are 
obtained.

\subsection{Uniqueness of ${\cal N}=2$ pure sugra}
\label{subsec:ConsistentFlat}

In this subsection, the single Latin index labelling the Maxwell gauge field will be omitted. We introduce no Majorana 
spin-$\tfrac{1}{2}$ field $\chi\,$. 
The upper-case Greek indices labelling the various gravitini run from 1 to 2
and are raised and lowered with the Euclidean metric in the internal space of 
the gravitini. 

\paragraph{First order deformation.}

In order to solve the last equation \eqref{ga20} of the descent 
for 
the algebra-deforming cubic candidates denoted by $a_2\,$,  
we have to classify all the $a_2 \in H(\gamma)$ that are cubic, 
i.e., of the type ``$C^\star CC$'' since, in \eqref{eq:masteraction}, 
the term $\tfrac{1}{2}\,C^\star_\alpha
  {f_0}^\alpha{}_{\beta\gamma} C^\beta C^\gamma$ encodes the 
  information about the gauge algebra. 
By using the result on the cohomology of $\gamma$ recalled in 
\eqref{cohomologyMink},
all the possible $a_2$'s that are parity and Poincar\'e invariant 
are listed as follows:
\begin{align}
a_2^{(1)} &=\, \xi^{\star\mu} \xi^\nu \partial_{[ \mu} \xi_{\nu ]}\;,\quad
a_2^{(2)} \,=\, \tfrac{1}{4}\, k^{(2)}_{\Delta\Omega}\, \xi^{\star\mu} 
\bar{\zeta}^{(\Delta} \gamma_\mu \zeta^{\Omega)},~~~
k^{(2)}_{\Delta\Omega} = k^{(2)}_{(\Delta\Omega)}\;,
\label{eq:3.1}\\
a^{(3)}_2 &=\, \xi^{\star\mu} \xi_\mu \mathcal{C}\;,\quad 
a_2^{(4)} \,=\, k^{(4)}_{\Delta\Omega}\, \mathcal{C}^\star \bar{\zeta}^{[\Delta} 
\zeta^{\Omega]},~~~k^{(4)}_{\Delta\Omega} = k^{(4)}_{[\Delta\Omega]}\;,
\quad a_2^{(5)} \,=\, k^{(5)}_{\Delta\Omega}\, \zeta^{\star\Delta} \zeta^{\Omega} \mathcal{C}\;,
\label{eq:3.2}\\
a_2^{(6)} &=\, k_{\Delta\Omega}^{(6)} \,\zeta^{\star\Delta} \gamma^{\mu} \zeta^{\Omega} \xi_\mu \;,\quad
a_2^{(7)} \,=\, 
k^{(7)}_{\Delta\Omega}\, \zeta^{\star\Delta} \gamma^{\mu\nu}
\zeta^{\Omega} \partial_{[\mu} \xi_{\nu]} \;.
\label{eq:3.3}
\end{align}
Thus the most general gauge-algebra deformation at first order, 
is a priori a linear combination of these seven candidates with the 
same number of associated infinitesimal deformation parameters.
\vspace{.3cm}

The next step amounts to solving equation \eqref{da2ga1} for the $a_1$'s.
A direct computation gives
\begin{align}
a_1^{(1)} &=\, h^{\star\mu\nu} \left[ \partial_\mu {\xi}^\sigma 
h_{\nu\sigma} - {\xi}^\sigma (\partial_\mu h_{\nu\sigma} 
- \partial_\sigma h_{\mu\nu}) \right]\;,\quad
a_1^{(2)} \,=\, - k^{(2)}_{\Delta\Omega}\, h^{\star\mu\nu} \bar{\psi}_{\mu}^\Delta 
\gamma_{\nu} {\zeta}^{\Omega}\;,\label{a1a}\\
a_1^{(3)} &=\, h^{\star\mu\nu} \left( 2 A_\mu \xi_\nu - h_{\mu\nu} \mathcal{C} 
\right)\;,\quad
a_1^{(4)} \,=\, - 2 k_{\Delta\Omega}^{(4)}\, A^{\star\mu} \bar{\psi}^{\Delta}_\mu 
\zeta^\Omega\;,\label{a1b}\\
a_1^{(5)} &=\, k^{(5)}_{\Delta\Omega}\, \psi^{\star\mu\Delta} 
\left( \psi_{\mu}^\Omega \,\mathcal{C} - \zeta^{\Omega}\, A_\mu \right)\;,
\label{eq:a15}\\
a_1^{(7)} &=\, \psi^{\star\Delta\rho} 
k_{\Delta\Omega}^{(7)} \, \gamma^{\mu\nu} \,
(\psi_{\rho}^\Omega \partial_{[\mu} \xi_{\nu]} 
- \zeta^\Omega \partial_{[\mu}h_{\nu]\rho})\;.\label{a1c}
\end{align}
Equation \eqref{da2ga1} has no nontrivial solution for $a_1^{(6)}$ as a function of 
$a_2 = a_2^{(6)}\,$, therefore we have to set to zero the coefficient in front of it, 
in the linear combination of the seven $a_2$'s.
Indeed, we find
\begin{equation}
\delta a_2^{(6)} = \partial_\mu j^\mu_1 - \gamma \left( k^{(6)}_{\Delta\Omega} \psi^{\star\Delta\rho} \gamma_\mu \psi_\rho^\Omega \xi^\mu - \frac{1}{2} k^{(6)}_{\Delta\Omega} \psi^{\star\Delta\rho} \gamma^\mu \zeta^\Omega h_{\mu\rho} \right) + k^{(6)}_{\Delta\Omega} \psi^{\star\Delta\rho} \gamma^\mu \zeta^\Omega \partial_{[\rho} \xi_{\mu]}
\end{equation}
and observe that the last term in the above equation is in $H(\gamma)\,$ 
and therefore represents an obstruction to finding a solution for $a^{(6)}_1\,$.
Note that the solution for the six elements $a_1$ found above in 
\eqref{a1a}--\eqref{a1c} can be supplemented by the general solution 
of the homogeneous equation 
\begin{align}
    \gamma  a_1= 0 \;,\label{ga10}
\end{align}
where a total derivative term in the right-hand side has been absorbed 
by a trivial redefinition, as done e.g. in \cite{Boulanger:2000rq,Boulanger:2001wq}.
In the following, we will use the notation  
$\hat{a}_1$ to indicate antifield number 1 cocycles at ghost number zero
that obey \eqref{ga10} and can be lifted to an $a_0$, i.e., such that 
$\delta \hat{a}_1 + \gamma a_0 = \partial_\mu \beta^\mu\,$ is true for some 
$a_0$ and $\beta^\mu\,$.
In other words such an $\hat{a}_1$ provides a solution of the descent equations of subsection 2.1 that does not contain an $a_2$ term, i.e, that does not deform the gauge algebra.
On the other hand, 
we will use the notation $\tilde{a}_1$ for antifield number 1 cocycles at 
ghost number zero that obey \eqref{ga10} and that are necessary to add to 
a candidate $a_1$ that comes from a corresponding $a_2\,$, when an obstruction 
arises in lifting $a_1$ alone.

\vspace{.3cm}

We now turn to the resolution of equation \eqref{da1ga0} for the 
candidate cubic vertices associated with the linear combination of the above 
six $a_1$'s. The resolution of equation \eqref{da1ga0} 
with $a_1 = a_1^{(1)}$ as source can be done separately and gives the cubic 
part of the Einstein-Hilbert action, as was done in \cite{Boulanger:2000rq}.
The resolution of \eqref{da1ga0} with $a_1^{(2)}$ and $a_1^{(7)}$ as sources 
has to be done concurrently, as these two terms talk to each other 
in the solution for the vertex and do not mix with the others. 
It turns out, as was observed in \cite{Boulanger:2001wq}, that
in order to find a vertex $a_0$ from the two sources $a_1^{(2)}$ and $a_1^{(7)}\,$,
one must add to them the following solution of the homogeneous equation \eqref{ga10}:
\begin{equation}
\tilde{a}_1^{(2-7)} = 
\psi^{\star\Delta\mu} \tilde{k}_{\Delta\Omega}\, \partial_{[\mu} \psi_{\nu]}^\Omega   \xi^\nu\;.\label{eq:3.10}
\end{equation}
Then, by use of $\gamma(\partial_{[\nu} h_{\rho]\mu}) = \partial_{\mu} \partial_{[\nu} \xi_{\rho]}$ and the relation \eqref{gamma3gamma2}, equation \eqref{da1ga0} 
admits a nontrivial solution if and only if
\begin{equation}
k^{(2)}_{\Delta\Omega} = k^{(7)}_{\Delta\Omega} = -\tfrac{1}{8} \tilde{k}_{\Delta\Omega} \eqqcolon k_{\Delta\Omega}\,,
\end{equation}
implying that $k^{(7)}_{\Delta\Omega}$ and $\tilde{k}_{\Delta\Omega}$
both inherit the symmetry of $k^{(2)}_{\Delta\Omega}\,$: 
$k_{\Delta\Omega} = k_{(\Delta\Omega)}\,$.
The solution for the corresponding cubic vertices explicitly reads
\begin{align}
a_0^{(2-7)} &= \bar{\psi}_\mu^\Delta \gamma^{\mu\nu\rho} \left( \sigma^{\alpha\beta} 
\partial_{[\alpha} h_{\beta]\nu} \right) \psi_\rho^\Omega k_{\Delta\Omega} 
- h \bar{\psi}_\mu^\Delta \gamma^{\mu\nu\rho} \partial_\nu \psi_\rho^\Omega 
\,k_{\Delta\Omega} \notag\\
\label{CubicExpansionLRS}
&~~~+ h_{\alpha\beta} \bar{\psi}^{\alpha\Delta} \gamma^{\beta\nu\rho} \partial_\nu 
\psi_\rho^\Omega k_{\Delta\Omega} + h_{\alpha\beta} \bar{\psi}_\mu^\Delta 
\gamma^{\mu\alpha\nu} \partial^\beta \psi_\rho^\Omega k_{\Delta\Omega} 
+ h_{\alpha\beta} \bar{\psi}_\mu^\Delta \gamma^{\mu\nu\alpha} \partial_\nu 
\psi^{\beta\Omega} k_{\Delta\Omega}\;.
\end{align}
Continuing with the solution of \eqref{da1ga0} starting from the other sources given 
in \eqref{a1a}--\eqref{a1c}, we see that the source $a_1^{(3)}$ presents an 
obstruction to admitting a  corresponding cubic vertex $a_0^{(3)}$. 
Indeed, a potential solution $a_0^{(3)}$ has to be linear in $A_\mu\,$, 
quadratic in $h_{\mu\nu}$ and must possess one derivative.
All the Poincar\'e-invariant possibilities are covered by 
$a_0^{(3)} = \sum\limits_{i=1}^6 x_i b^i,~x_i \in \mathbb{R}\,$, 
where
\begin{align}
b^1 &= A^\mu \partial_\mu h^{\alpha\beta} h_{\alpha\beta},~~~~~~~b^2 =A^\mu \partial_\mu h~ h,~~~~~~~b^3 =A^\mu \partial_\alpha h_{\mu\beta} h^{\alpha\beta},\notag\\
b^4 &=A^\mu \partial^\alpha h_{\mu\alpha}h,~~~~~~~~~~b^5 =A_\mu \partial_\alpha h~ h^{\mu\alpha},~~~~b^6 =A^\mu \partial_\beta h^{\beta\alpha} h_{\alpha\mu}\;.
\end{align}
A direct computation shows that there is no possible choice for the coefficients 
$x_i$ that can solve \eqref{da1ga0}. 
To cure this situation one cannot resort to an 
$\tilde{a}_1^{(3)}$ solution of the 
homogeneous equation \eqref{ga10}, because an element of $H(\gamma)$ that has 
the right structure in terms of fields, ghosts and antifields 
will necessarily bring too many derivatives and therefore 
cannot mix up with $\delta a_1^{(3)}$ to solve \eqref{da1ga0} with 
$a_1^{(3)}+\tilde{a}_1^{(3)}$ as source for $a_0^{(3)}\,$.
\vspace{.3cm}

We now continue the process and consider $a_1^{(4)}$ as source for a potential 
vertex $a_0^{(4)}\,$. By adding
\begin{align}
    \tilde{a}_1^{(4)} \coloneqq \tilde{t}_{\Delta\Omega}\, \psi^{\star\Delta}_{\phantom{\star}\sigma} F_{\mu\nu} \gamma^{\mu\nu} \gamma^\sigma \zeta^\Omega
    \label{tildea4}
\end{align}
to $a_1^{(4)}\,$, 
it becomes possible to solve \eqref{da1ga0} for an $a_0^{(4)}\,$, if and only if
\begin{equation}
\tilde{t}_{\Delta\Omega} = \tfrac{1}{2} k^{(4)}_{\Delta\Omega} 
\eqqcolon t_{\Delta\Omega}\;,
\end{equation}
therefore we have that $t_{\Delta\Omega} = t_{[\Delta\Omega]}$ must be 
antisymmetric because $k^{(4)}_{\Delta\Omega}$ has this property. 
In particular, this proves the impossibility to couple, minimally or not,
a single Majorana spin-$\tfrac{3}{2}$ gauge field to 
electromagnetism, as was announced in the abstract and in the introduction.
Furthermore, since the upper-case Greek indices run over only two values, we 
take
\begin{equation}
t_{\Delta\Omega} = \epsilon_{\Delta\Omega}\;,
\end{equation}
the $\mathfrak{sp}_2$-invariant antisymmetric symbol, 
up to a  coefficient that we will insert when considering a 
linear combination of all the $a_1$'s.
By using the relations \eqref{gamma3gamma1gamma2}, \eqref{gamma2gamma1gamma3gamma1}, \eqref{commgamma1gamma2gamma1} and \eqref{gamma2gamma2} the cubic vertex $a_0^{(4)}$ that we obtain reads
\begin{equation}
a_0^{(4)} = \epsilon_{\Delta\Omega} \bar\psi_\mu^\Delta \left( F^{\mu\nu} \Id -i \gamma^5 (*F)^{\mu\nu} \right) \psi_\nu^\Omega\;,\qquad 
(*F)^{\mu\nu}\coloneqq\tfrac{1}{2}\,\epsilon^{\mu\nu\rho\sigma}\,F_{\rho\sigma}\;,
\label{eq:3.17}
\end{equation}
where $\epsilon^{\mu\nu\rho\sigma}$ is the tensorial quantity such that 
$\epsilon^{0123}=1$ in Cartesian coordinates.  
This cubic vertex is part of the interactions of the  
$\mathcal{N}_4 = 2$ pure supergravity Lagrangian of \cite{Ferrara:1976fu} 
and is a Pauli dipolar coupling. 
\vspace{.3cm}

Finally, when trying to solve \eqref{da1ga0} for a possible vertex $a_0^{(5)}\,$, 
with $a_1^{(5)}$ as the source, we find an obstruction. 
This is the well-known problem \cite{Johnson:1960vt,Velo:1969bt} of the minimal coupling 
of gravitini to electromagnetism around flat space.
We defer the detailed cohomological derivation of this no-go result 
to the next subsection.
\vspace{.3cm}

Before doing this, we want to mention that the coupling of the vector field 
to gravity is produced by the following 
$\hat{a}_1^{_{G1}}\,$, solution of \eqref{ga10}:
\begin{equation}
\hat{a}_1^{_{G1}} = - A^{\star\mu} F_{\mu\nu} \xi^\nu\;.\label{eq:3.18}
\end{equation}
This cocycle of $\gamma$ gives rise to the following vertex
\begin{equation}
a_0^{_{G1}} = -\tfrac{1}{8} h F^{\mu\nu} F_{\mu\nu} -\tfrac{1}{2} F^{\mu\sigma} F_{\sigma}^{\phantom{\sigma}\nu}\;.\label{eq:3.19}
\end{equation}
This vertex does not deform the gauge algebra as it arises from the 
candidate $\hat{a}_1^{_{G1}}\,$, solution of \eqref{ga10} without $\delta a_2$
source term.

\paragraph{Summary at first order.}
We gather the various results we obtained so far at first order in 
the deformation parameters:
\begin{align}
a_2 &= \kappa\, a_2^{(1)} + \alpha \left(a_2^{(2)} + a_2^{(7)}\right) + y \,a_2^{(4)},\\
a_1 &= \kappa\, a_1^{(1)} + \alpha \left(a_1^{(2)} + a_1^{(7)} + \tilde{a}_1^{(2-7)}\right) 
+ y (a_1^{(4)} + \tilde{a}_1^{(4)}) + \beta \,\hat{a}_1^{_{G1}},\\
a_0 &= \kappa\, a_0^{(1)} + \alpha \,a_0^{(2-7)} + y \,a_0^{(4)} + \beta\, a_0^{_{G1}}\;,
\label{eq:3.22}
\end{align}
where $\kappa$, $\alpha$, $\beta$ and $y$ are (yet unrelated) 
infinitesimal deformation parameters that constitute the set of parameters 
that we collectively denoted by $g$ above.
The cocycles $a_2^{(1)}\,$, $a_2^{(2)}\,$,  $a_2^{(7)}\,$ and 
$a_2^{(4)}$ are given in \eqref{eq:3.1} -- \eqref{eq:3.3}. They encode 
the first-order deformations of the gauge algebra and have an antifield 
number two: $antifd(a_2) = 2\,$. 
For example, $\kappa\, a_2^{(1)} = \kappa\, \xi^{\star\mu} \xi^\nu \partial_{[ \mu} \xi_{\nu ]}\,$ gives the Lie algebra of vector fields equipped with the Lie bracket. 
Indeed, adding to it the trivial piece 
$\gamma (\tfrac{\kappa}{2}\,\xi^{\star\mu} \xi^\nu h_{\mu\nu})$ gives an equivalent 
element $a_2\coloneqq \kappa \,\xi^{\star\mu} \xi^\nu \partial_{\nu} \xi_{\mu}$
from which we read the result announced and produced in \cite{Boulanger:2001wq}. 
The cocycles $a_1^{(1)}\,$,  $a_1^{(2)}\,$, $a_1^{(7)}\,$ and 
$a_1^{(4)}$ are given in \eqref{a1a} --\eqref{a1c}, while 
$\tilde{a}_1^{(2-7)}\,$, $\tilde{a}_1^{(4)}$ and $\hat{a}_1^{_{G1}}$
are respectively given in  \eqref{eq:3.10}, \eqref{tildea4} and 
\eqref{eq:3.18}. 
These cocycles $a_1$'s encode the deformations of the gauge 
transformations at first order and have $antifd = 1\,$.
Finally, the corresponding vertices are given in \eqref{CubicExpansionLRS}, 
\eqref{eq:3.17} and \eqref{eq:3.19}. 
They have zero antifield number: $antifd(a_0) = 0\,$.

\paragraph{Second order deformations.}

We are going to investigate whether the first order deformations can 
allow for second order extensions. 
In order to do this, we first solve \eqref{a2a2} for a deformation of the 
gauge algebra, $b_2\,$, to second order in the parameters.
By means of $\gamma(\partial_{[\nu} h_{\rho]\mu}) = \partial_{\mu} \partial_{[\nu} \xi_{\rho]}$, \eqref{gamma1gamma2gamma3gamma1}, \eqref{commgamma1gamma2gamma1}, \eqref{gamma2gamma2} and the Fierz identity
\begin{equation}
    k_{\Delta\Sigma} k_{\Omega\Gamma} (\zeta^{\star\Delta} \zeta^\Omega) (\bar{\zeta}^\Sigma \zeta^\Gamma) = -\tfrac{1}{4} k_{\Delta\Sigma} k_{\Omega\Gamma} (\zeta^{\star\Delta} \gamma_\mu \zeta^\Sigma) (\bar{\zeta}^\Omega \gamma^\mu \zeta^\Gamma) + \tfrac{1}{8} k_{\Delta\Sigma} k_{\Omega\Gamma}  (\zeta^{\star\Delta} \gamma_{\mu\nu} \zeta^\Sigma) (\bar{\zeta}^\Omega \gamma^{\mu\nu} \zeta^\Gamma)\,,
\end{equation}
we find that all obstructions to finding $b_2$ are cancelled if and only if
\begin{equation}
\alpha = \frac{\kappa}{4}\,,\quad 
k^\Delta{}_{\Sigma} k^{\Sigma}_{\phantom{\Sigma}\Omega} = k^\Delta{}_{\Omega}\;.
\end{equation}
In this section we suppose that the two spin-3/2 gauge fields couple to gravity. 
It implies that the matrix of coefficients $k^\Delta{}_{\Omega}$ is of maximal rank. 
In this case the above constraint is solved if and only if 
$k^\Delta{}_{\Omega} = \delta^\Delta_\Omega\,$. 
We will be more general and not do this assumption in  
subsection \ref{subsection:3.4} where the number of spin-3/2 
gauge fields can remain arbitrary.
The solution is given by
\begin{align}
b_2 &= \tfrac{\kappa^2}{4}\;\xi^{\star\alpha}\, \xi_\mu
    \Big[ h_{\alpha\nu}\, \partial^{[\mu}\xi^{\nu]} - h^{\mu\nu}\, \partial_{[\alpha}\xi_{\nu]} 
    - \partial^{[\mu}h^{\nu]}{}_\alpha\, \xi_{\nu} \Big] + \tfrac{\kappa^2}{8}\; \xi^\star_\mu\, \xi_\nu\, \bar{\zeta}^\Delta \gamma^{[\mu} \psi^{\nu]}_\Delta\notag\\
&~~~+ \tfrac{\kappa^2}{4}\; \zeta^\star_\Delta \gamma_{\mu\nu} \zeta^\Delta \left( \tfrac{1}{2}\, h^\mu_{\phantom{\mu}\sigma} \partial^{[\nu} \xi^{\sigma]} + \xi_\sigma \partial^{[\sigma} h^{\nu]\mu} \right) + \tfrac{\kappa^2}{32} \left( \zeta^\star_\Delta \gamma^{\mu\nu} \zeta^\Delta \right) \left( \bar{\zeta}_\Omega \gamma_\mu \psi_{\nu}^{\Omega} \right)\;.
\end{align}
It encodes the gauge algebra to second order in the deformation.
From now on there are 3 independent infinitesimal deformation parameters:
$\kappa\,$, $y$ and $\beta\,$.
The pure $\mathcal{N}_4 = 2$ supergravity theory without cosmological constant 
possesses a single dimensionful constant, 
namely the Newton constant, therefore we must expect that the equations 
\eqref{sgammaa1} and \eqref{db1ga0a1a0} will impose constraints on the 
remaining constants, so that $\beta$ and $y\,$, say, will
depend on $\kappa\,$. 
Let us proceed with equation \eqref{sgammaa1} and consider, therein, 
all the terms that have the schematic structure 
``$A^\star F \xi \xi\,$'' $\in H(\gamma)$ 
and therefore represent an obstruction to finding $b_1\,$, 
the deformations of the gauge 
transformation at second order in perturbation. 
We will give later --- in \eqref{alpha1kappa} where $\beta$ is called 
$\alpha_1$ --- the details of this 
computation when the background is $AdS_4$ and with an arbitrary number
of Rarita-Schwinger and  Maxwell gauge fields. 
It turns out that the only way to set to zero the coefficient 
in front of this obstruction is by imposing the relation
\begin{equation}
\beta = \kappa\;.
\end{equation}
For what concerns the analysis of the terms of the schematic form 
$A^\star F \zeta \zeta$ $\in H(\gamma)\,$ --- we defer the treatment to \eqref{AstarFzetazeta}, in 
$AdS_4$ background and for an arbitrary number ${\cal N}_4$
of Rarita-Schwinger gauge fields and $n_v$ Maxwell fields ---, 
we have to impose
\begin{equation}
y^2 = \frac{\kappa^2}{32}\;
\end{equation}
in order to cancel the coefficient in front of the obstruction. 
This leaves us with only one deformation parameter $\kappa\,$.
Let us show that we reproduced the pure $\mathcal{N}_4 = 2$ supergravity 
theory of \cite{Ferrara:1976fu}.

\paragraph{Analysis of the deformation.}

Omitting the terms quartic in the spinors, 
the Lagrangian density $\cal L$ in \cite{Ferrara:1976fu} is given by 
\begin{equation}
{\cal L} = \frac{\sqrt{-g}}{\kappa^2}\,\left[ L^{EH} 
- \tfrac{\kappa^2}{2} 
\bar{\psi}^{\Delta}_\mu \gamma^{\mu\nu\rho} 
\mathcal{D}_\nu \psi^\Omega_{\rho} \, \delta_{\Delta\Omega} 
- \tfrac{\kappa^2}{4} g^{\mu\rho} 
g^{\nu\sigma} F_{\mu\nu} F_{\rho\sigma} \pm \tfrac{\kappa^3}{4\sqrt{2}} \psi_\mu^\Delta 
\left( F^{\mu\nu} \Id -i \gamma^5 (*F)^{\mu\nu} \right) \psi_\nu^\Omega 
\epsilon_{\Delta\Omega}\right] \;,
\label{LagrangianFvN}
\end{equation}
where $\mathcal{D}_\mu \psi_\nu = \partial_\mu \psi_\nu + \tfrac{1}{4} \omega_\mu{}^{ab} \gamma_{ab}\,\psi_\nu$ is the Lorentz-covariant derivative on spinors.
Since we study the deformations of a free theory around 
flat spacetime, 
we have\footnote{The symbol $\kappa$ is chosen so that it corresponds to 
the dimensionful constant (related to the Newton constant by 
$\kappa^2 = \frac{32\pi G}{c^4}$) that appears in the perturbative 
expansion of the metric.}
\begin{align}
g_{\mu\nu} &\equiv e^a_{\phantom{a}\mu} e_{a\nu} = \eta_{\mu\nu} + \kappa \,h_{\mu\nu}\;,
\quad 
g^{\mu\nu} \,=\, \eta^{\mu\nu} - \kappa \,h^{\mu\nu} + {\cal O}(h^2),\\
e^a_{\phantom{a}\mu} &= \delta^a_\mu + \frac{\kappa}{2}\; h^a_{\phantom{a}\mu}\;,\quad 
e_a^{\phantom{a}\mu} \,=\, \delta_a^\mu - \frac{\kappa}{2}\; h_a^{\phantom{a}\mu}
+ {\cal O}(h^2) \;,
\quad e \,\equiv\, \sqrt{-g} = 1+ \frac{\kappa}{2} \;h+ {\cal O}(h^2)\;.
\end{align}
Let us show that the cubic vertices \eqref{eq:3.22} that 
we have derived are the cubic part of the $\mathcal{N}_4 = 2$ supergravity 
Lagrangian density $\cal L\,$. 
Firstly, the last term in \eqref{LagrangianFvN} is directly given 
by $a_0^{(4)}\,$ of \eqref{eq:3.17}. 
Secondly, the expansion to third order of the third term in \eqref{LagrangianFvN}
is given by $a_0^{_{G1}}\,$. 
Finally, the cubic expansion of the second term in \eqref{LagrangianFvN} 
correctly gives $a_0^{(2-7)}\,$ in \eqref{CubicExpansionLRS}. 
Actually, the first term of \eqref{CubicExpansionLRS} 
gives the cubic part of $-\tfrac{1}{4}\,\bar{\psi}^{\Delta}_\mu \,
\delta^{\mu}_a \delta^{\nu}_b \delta^{\rho}_c\,\gamma^{abc} \,\omega_\nu{}^{pq} \,\sigma_{pq} \psi^\Omega_{\rho}\,\delta_{\Delta\Omega}\,$,  
while the others represent the cubic part of 
$e\,\bar{\psi}^{\Delta}_\mu\, \gamma^{abc} \,\partial_\nu \psi^\Omega_{\rho}\,  e_a^{\phantom{a}\mu} e_b^{\phantom{b}\nu} e_c^{\phantom{c}\rho} \,\delta_{\Delta\Omega}\,$.
\vspace{.3cm}

We recall that the parts of the susy gauge transformations of the 
$\mathcal{N}_4 = 2$ supergravity theory that are linear in the fields 
are given by
\begin{align}
\delta^{(1)}_{\epsilon} A_\mu ~&\propto~ \bar{\epsilon}^\Delta \psi^\Omega_\mu \epsilon_{\Delta\Omega}\;,\\
\delta^{(1)}_{\epsilon} e^a_{\phantom{a}\mu}~&\propto ~\bar{\epsilon}^\Delta \gamma^a \psi_\mu^\Omega \delta_{\Delta\Omega}\;,\\
\delta^{(1)}_{\epsilon} \psi^\Delta_\mu ~&\propto~ 
\mathcal{D}_\mu \epsilon^\Delta + \epsilon^{\Delta\Omega}\,\left( F_{\mu\nu} \gamma^\nu 
+ i (*F)_{\mu\nu} \gamma^\nu \gamma^5 \right) \epsilon_\Omega\;.
\end{align}
The susy transformations of $A_\mu$ and $h_{\mu\nu}$ are directly given 
by $a_1^{(2)}$ and $a_1^{(4)}\,$ in \eqref{a1a} and \eqref{a1b}. 
By use of \eqref{gamma2gamma1gamma3gamma1} and 
\eqref{gamma3epsgamma5} one can rewrite the last part of the susy 
gauge transformation of the gravitini as in $\tilde{a}_1^{(4)}$ \eqref{tildea4}, 
the first part being written in the second term of $a_1^{(7)}$ (see \eqref{a1c}) 
because $\stackrel{(1)}{\omega}_{\mu}{}^{\alpha\beta} = - \partial^{[\alpha} h^{\beta]}{}_\mu$ is the linearised spin connection in the Lorentz gauge where 
the antisymmetric part of the vierbein components is zero.
\vspace{.3cm}

The rest of the gauge transformations is not written in \cite{Ferrara:1976fu} 
because they are just the transformations under diffeomorphisms. 
It should also be noted that, without cosmological constant, there is no 
deformation of the gauge transformations involving the $U(1)$ parameter.
In other words, the gravitini remain uncharged under $U(1)\,$. 
The consistent minimal coupling of the gravitini to electromagnetism is the subject of the 
next subsection.

\subsection{Obstruction to the $U(1)$ minimal coupling.}
\label{subsec:Obstru}

This subsection is devoted to the obstruction coming from the resolution 
of \eqref{da1ga0} with source term $a_1^{(5)}$ \eqref{eq:a15}. 
Acting with the Koszul-Tate differential on $a_1^{(5)}$ yields
\begin{equation}
\label{3.34}
\delta a_1^{(5)} = \partial_\mu j^\mu - \gamma \left( \tfrac{1}{2} k^{(5)}_{\Delta\Omega} \bar{\psi}^\Delta_\rho \gamma^{\mu\nu\rho} \psi_\nu^\Omega A_\mu \right) + \underbrace{\tfrac{1}{2} k^{(5)}_{\Delta\Omega} F_{\mu\nu} \bar{\psi}^\Delta_\rho \gamma^{\mu\nu\rho} \zeta^\Omega}_{Obstruction}\;.
\end{equation}
The last term is an obstruction to finding a cubic vertex $a_0^{(5)}$ 
in the resolution of \eqref{da1ga0}, because it cannot be written 
as a $\gamma\,$-exact term modulo $\partial_\mu j^{\mu}\,$, 
for both the antisymmetric and the symmetric parts of $k^{(5)}_{\Delta\Omega}\,$.
This can be seen by taking the Euler-Lagrange derivative of it with respect 
with $A_\mu$ and checking that it gives an element in $H(\gamma)\,$.
The only way to cancel this obstruction without adding extra deformation 
is by choosing $k^{(5)}_{\Delta\Omega}$ to be identically zero. 
Treating the resolution of $a_2^{(5)}$ together with another $a_2$ or adding an 
$\tilde{a}_1^{(5)} \in H(\gamma)$ does not help because 
the Koszul-Tate differential $\delta$ acting 
on them gives at least two derivatives while the obstruction only has 
one derivative. 
\vspace{.3cm}

The nontrivial way to cancel this obstruction,  
thereby saving the minimal coupling in a Poincar\'e-invariant 
deformation procedure around flat background, is by introducing an 
infinitesimal deformation that is \emph{linear} in the dynamical 
fields. 
The only such Poincar\'e-invariant deformation is 
\begin{equation}
{W}^{lin.}_1 = \int d^4x~ (-2 \tilde\lambda^2\, h)\;.
\label{3.35}
\end{equation}
It is a first order deformation of the master equation 
because the BRST differential acting on $h=\eta^{\mu\nu}h_{\mu\nu}$ 
gives a total derivative \cite{Boulanger:2000rq}.
It is easy to see that there is no other Poincar\'e-invariant, 
infinitesimal deformation that is linear in the fields of the 
spectrum at hand.  
If one accepts to break Poincar\'e symmetry by explicitly 
introducing Cartesian coordinates $x^\mu$ dependence into 
the higher-order deformations, the above deformation 
can be seen to trigger a change of background, leading 
from flat to $(A)dS_4$ background\footnote{The $AdS_4$ background
permits unitary representation for the spinors. 
One way to accommodate the $dS_4$ background is by changing 
the reality condition on the spinors, but as explained in 
\cite{Pilch:1984aw}, 
in the interacting theory the sign in front of the Maxwell action 
will have to flip in order to keep the vector fields real, 
leading to a classical ghost. Alternatives to supergravity models 
in de Sitter background can be found 
in \cite{Bergshoeff:2015tra} and refs. therein.}. 
Note that, with the first-order infinitesimal deformation \eqref{3.35},
linear in the weak field, the degree of polynomiality of the 
deformation is not related to the degree of perturbation by a shift of 2. 
For example, around flat space, 
with the infinitesimal cosmological constant deformation \eqref{3.35} being 
introduced, cubic vertices may appear at second order in deformation, 
instead of appearing at first order. 
\vspace{.3cm}

Sticking to Poincar\'e invariance for the moment 
and in order for the 
deformation to be consistent at second order 
in the presence of real Rarita-Schwinger gauge fields, 
one also needs \cite{Boulanger:2001wq} to add a mass term for the 
massless spin-$\tfrac{3}{2}$ fields,  
which translates into the following additional 
infinitesimal deformations, this time quadratic in the fields:
\begin{align}
a_1^{m} &= -\tfrac{m}{2}\; \psi^{\star\mu}_{\phantom{\star}\Delta} \gamma_\mu \zeta^\Delta\;,\\
a_0^m &= -\tfrac{m}{2} \;\bar{\psi}^{\mu\Delta} \gamma_{\mu\nu} \psi^\nu_\Delta\;,
\label{3.37}
\end{align}
the mass being related to the cosmological constant 
that has to be negative for Majorana spinors, as was re-derived in 
\cite{Boulanger:2001wq}, consistently with the original 
finding of \cite{Townsend:1977qa}.
\vspace{.3cm}

As explained in the Introduction, instead of proceeding with the 
study of the Poincar\'e-invariant deformations around flat space 
that include the new cosmological pieces given above, a procedure 
that can certainly be pursued but has the aesthetic inconvenience 
of having first-order deformations mixing up expressions of first, 
second and third polynomial orders in the weak fields, 
the strategy we prefer (and that leads to the same results) 
is to start the deformation procedure from the solution of the 
classical master equation around $AdS_4$ with finite (negative) 
cosmological constant, asking for $SO(2,3)$-invariant 
infinitesimal deformations. 
Since in all cases we see that we have to introduce gravity and 
the diffeomorphism algebra for consistency of the couplings,
thereby leading to a background independent end result, 
the advantage with the latter approach  is that the minimal 
coupling terms will appear at first order in deformation around 
$AdS_4\,$, instead of appearing at second order in deformation around 
flat background.
The solution of the classical master equation associated 
with the free theory in $AdS_4$ 
background is given in \eqref{BVfunctionalAdS}.  
The flat limit $\lambda \longrightarrow 0$ of this functional is smooth 
and reproduces the solution of the master equation \eqref{masterequflatspace} for 
the free theory in Minkowski spacetime. 

\subsection{Deformations around AdS background}
\label{sec:ConsistentAdS}

In this section, we therefore start from the functional \eqref{BVfunctionalAdS} 
and compute its most general, infinitesimal cubic deformations that preserve the 
$SO(2,3)$ symmetries of the $AdS_4$ background.
As in the previous sections and following the general method of \cite{Barnich:1994mt}, 
we first classify all the possible deformations that deform the gauge algebra,
and for each of them, we determine whether they can be lifted to give 
deformations of the gauge transformations, and possibly a cubic vertex. 
Along that way, we also determine the deformations that do not alter 
the abelian gauge algebra of the free theory 
but that nevertheless deform the gauge transformations laws.   
The analyses of this section are performed for ${\mathcal{N}}_4$ arbitrary,
however, due to the spectrum of fields that we allow, consistent theories 
to all orders will only be obtained for the cases where ${\mathcal{N}}_4<4\,$. 
Some of the constraints we obtain (those that come 
from \eqref{a2a2}) remain unchanged whenever one introduces extra matter fields
in the form of scalars and Majorana spinors,
whereas other constraints will change when additional matter fields 
will be introduced. 
\vspace{.3cm}

We start with 
\begin{equation}
s {W}_1 = 0\;,
\end{equation}
set ${W}_1 = \int d^4x\, \sqrt{-\bar{g}} (a_0 + a_1 + a_2)$
and expand the above equation with respect to the antighost number:
\begin{align}
\delta a_1 + \gamma a_0 &= \partial_\mu j^\mu_0 = \sqrt{-\bar{g}} \;\nabla_\mu \tilde{j}^\mu_0\;,\label{eq:3.38}\\
\delta a_2 + \gamma a_1 &= \partial_\mu j^\mu_1 = \sqrt{-\bar{g}} \;\nabla_\mu \tilde{j}^\mu_1\;,\label{eq:3.39}\\
\gamma a_2 &= 0\label{eq:3.40}\;.
\end{align}
We search for solutions with non-trivial $a_2\,$ and 
therefore have to classify the possible cubic $a_2$'s belonging to 
the cohomological groups $H(\gamma)\,$ in pureghost number 2 and form degree 
zero. 
The classification of these terms is straigthforward and 
has already been presented in the flat space case in \eqref{eq:3.1}--\eqref{eq:3.3}. 
All we have to do is to covariantise these candidates to 
the $AdS_4$ background, adding mass-like terms when they are allowed, and 
adding the Yang--Mills deformation term 
$\tfrac{1}{2}\,{\cal C}^{\star}_a   f^a{}_{bc}\,{\cal C}^b{\cal C}^c$ since we now allow 
for an arbitrary number $n_v$ of free abelian vector fields. 
We present only those $a_2$'s that lead to corresponding $a_1$ and $a_0\,$, 
upon solving \eqref{eq:3.38}--\eqref{eq:3.40}, without encountering any 
obstruction. The final result reads 
$a_2 = \kappa\, a_2^{_{EH}} + \alpha_{_{{3}/{2}}}\, a_2^{susy} + y \,a_2^{cc} 
+ g_{_{YM}}\, a_2^{_{YM}}$ and generates the algebras of diffeomorphisms, 
local supersymmetries with ``central charges'' in $AdS_4$, and Yang--Mills, 
respectively. 
The constant parameters $\kappa$, $\alpha_{_{{3}/{2}}}\,$, 
$y$ and $g_{_{YM}}$ are infinitesimal deformation parameters that we choose 
in this paper not to absorb in the corresponding structure constants. 
We present the results of this classification in various 
paragraphs, with titles referring to the infinitesimal deformations of 
the gauge algebra for the various linear combinations of $a_2$'s.

\paragraph{Diffeomorphisms.}

The diffeomorphism algebra deformation, around $AdS_4$, is obtained from the 
following candidate:
\begin{equation}
a_2^{_{EH}} = \xi^{\star\mu} \xi^\nu \nabla_{[\mu} \xi_{\nu]}\;.
\end{equation}
This candidate can be lifted to a deformation of the gauge transformations
through
\begin{equation}
a_1^{_{EH}} = h^{\star\mu\nu} \left[ \nabla_\mu \xi^\sigma h_{\nu\sigma} - \xi^\sigma (\nabla_\mu h_{\nu\sigma} - \nabla_\sigma h_{\mu\nu}) \right]\;.
\end{equation}

Of course, as in \cite{Boulanger:2000rq} but this time in the $AdS_4$ background, 
one can make a trivial redefinition of $a_2^{_{EH}}$ in order to read in the 
corresponding $a_1$ the transformations of $h_{\mu\nu}$ given by
its  Lie derivative along $\xi^\mu\,$, the gauge parameters associated with 
infinitesimal diffeomorphisms.
Then we can lift $a_1^{_{EH}}$ to get the cubic vertex of the Einstein-Hilbert action with 
negative cosmological constant around $AdS_4$ in such a way that the 
cosmological constant terms appear explicitly as the cubic part of 
$\sqrt{-g}\,$.
The latter cubic terms can of course appear with different relative coefficients
compared to the mere expansion of $\sqrt{-g}\,$ to third order 
around $AdS_4$, 
the reason being the ambiguity in the cubic terms with two $AdS_4$-covariant derivatives, 
once one integrates by parts. We have fixed this freedom in integration by parts by 
requiring that the undifferentiated cubic terms exactly produce the cubic 
terms in the expansion of $\sqrt{-g}\,$ around $AdS_4$. In this way, we obtain the 
following cubic vertex:
\begin{align}
a_0^{_{EH}} &= \tfrac{1}{2} h^{\mu\nu} \nabla_\mu h_{\rho\sigma} \nabla_\nu h^{\rho\sigma} - \tfrac{1}{2} h^{\mu\nu} \nabla_\mu h \nabla_\nu h + h^{\mu\nu} \nabla_\mu h \nabla^\sigma h_{\sigma\nu} + h^{\mu\nu} \nabla_\mu h_{\nu\sigma} \nabla^\sigma h \notag\\
&~~~- h^{\mu\nu} \nabla_{\sigma} h_{\mu\nu} \nabla^\sigma h + \tfrac{1}{4} h \nabla_\mu h \nabla^\mu h + h^{\mu\nu} \nabla_\sigma h_{\mu\nu} \nabla_\rho h^{\rho\sigma} - \tfrac{1}{2} h \nabla_\mu h \nabla_\nu h^{\mu\nu} \notag\\
&~~~-2 h^\mu_{\phantom{\mu}\nu} \nabla_\mu h_{\rho\sigma} \nabla^\rho h^{\sigma\nu} - h^\mu_{\phantom{\mu}\nu} \nabla_\sigma h_{\rho\mu} \nabla^\rho h^{\sigma\nu} + h^\mu_{\phantom{\mu}\nu} \nabla_\sigma h_{\rho\mu} \nabla^\sigma h^{\rho\nu} + \tfrac{1}{2} h \nabla_\sigma h_{\mu\nu} \nabla^\mu h^{\nu\sigma}\notag\\
&~~~- \tfrac{1}{4} h \nabla_\sigma h_{\mu\nu} \nabla^\sigma h^{\mu\nu} + \lambda^2 \left( \tfrac{1}{2} hhh + 4 h_\mu^{\phantom{\mu}\nu} h_\nu^{\phantom{\nu}\rho} h_{\rho}^{\phantom{\rho}\mu} - 3 h h_{\mu\nu} h^{\mu\nu} \right)\;.
\end{align}

\paragraph{Supersymmetry algebra.}

The next gauge-algebra deformation candidate that can be 
lifted all  the way to a deformation of the gauge transformations and to a cubic vertex
reads
\begin{align}
\alpha_{_{{3}/{2}}}\,a_2^{susy} + y\, a_2^{cc-\Lambda} &= 
\alpha_{_{{3}/{2}}}\,\left( \tfrac{1}{4} \,\xi^{\star\mu} \bar{\zeta}_\Delta \gamma_\mu \zeta_{\Omega} k^{\Delta\Omega} + \zeta^{\star}_\Delta \gamma^{\mu\nu} \zeta_{\Omega} \nabla_{[\mu} \xi_{\nu]} k^{\Delta\Omega} - 2 \lambda \,\zeta^{\star}_\Delta \gamma^{\mu} \zeta_{\Omega} \xi_\mu k^{\Delta\Omega}\right) \notag\\
& \quad + y\, \left( \mathcal{C}^\star_a \bar{\zeta}^{[\Delta} \zeta^{\Omega]} t^a_{\phantom{a}\Delta\Omega} - 2 \lambda\, \zeta^{\star}_{\phantom{\star}\Delta} \zeta_{\Omega} \mathcal{C}^a t_a^{\phantom{a}\Delta\Omega}\right)\;,  
\quad t^a_{\phantom{a}\Delta\Omega} = - t^a_{\phantom{a}\Omega\Delta}\;.
\end{align}
The coefficient $-2\lambda$ in front of the third term above is fixed by 
requiring that the $a_{2}$ candidate can be lifted to an $a_{1}\,$. 
All other new --- as compared to the 
flat space case studied previously --- linear relations between the coefficients of the
 $a_{2}$'s  appear when lifting the $a_{1}$ to $a_{0}$. 
We can lift that linear combination to obtain the following infinitesimal 
gauge-transformation deformation $a_1\,$:
\begin{align}
\alpha_{_{{3}/{2}}}\,a_1^{susy}+ y\,a_1^{cc-\Lambda} = &
\; \alpha_{_{{3}/{2}}}\,\Big[
- h^{\star\mu\nu} \bar{\psi}_{\mu\Delta} \gamma_{\nu} \zeta_{\Omega} k^{\Delta\Omega} + \psi^{\star\rho}_{\phantom{\star}\Delta} \gamma_{\mu\nu} 
\psi_{\Omega\rho} \nabla^{[\mu} \xi^{\nu]} k^{\Delta\Omega} - 
\psi^{\star\rho}_{\phantom{\star}\Delta} \gamma^{\mu\nu} \zeta_{\Omega} 
\nabla_{[\mu}h_{\nu]\rho} k^{\Delta\Omega}\notag\\
&\qquad + \lambda \,\psi^{\star\mu}_{\phantom{\star}\Delta} \gamma^\nu \zeta_{\Omega} h_{\mu\nu} 
k^{\Delta\Omega} - 2 \lambda \,\psi^{\star\mu}_{\phantom{\star}\Delta} \gamma_\nu 
\psi_{\Omega\mu} \xi^\nu k^{\Delta\Omega}\Big]\nonumber \\
  & + \,y\,\Big[ - 2 A^{\star\mu}_{\phantom{\star}a} \bar{\psi}^{\Delta}_\mu \zeta^\Omega t^a_{\phantom{a}\Delta\Omega} - 2\lambda \psi^{\star\mu}_{\phantom{\star}\Delta} \left( \psi_{\mu\Omega} \mathcal{C}^a - \zeta_{\Omega} A_\mu^a \right) t_a^{\phantom{a}\Delta\Omega} \Big]\;.
  \label{eq:3.47}
\end{align}
This term can be lifted to get a vertex $a_0$ only provided we add to it an 
appropriate cocycle $\tilde{a}_1$ of $H(\gamma)\,$, that is to say, 
an appropriate solution to the homogeneous equation $\gamma \tilde{a}_1 = 0\,$:
\begin{equation}
\alpha_{_{{3}/{2}}}\,\tilde{a}_1^{_{G3/2}} + y\,\tilde{a}_1^{cc} =
-8\alpha_{_{{3}/{2}}}\, \psi^{\star\mu}_{\phantom{\star}\Delta} \Psi_{\mu\nu\Omega} 
\xi^\nu k^{\Delta\Omega} -\tfrac{1}{2} \,y\,\psi^{\star}_{\mu\Delta} F^a_{\rho\sigma} \gamma^{\rho\sigma} \gamma^\mu \zeta_{\Omega} t_a^{\phantom{a}\Delta\Omega} \;.
\end{equation}
Only after the addition of this term to \eqref{eq:3.47} can we 
get from the equation \eqref{eq:3.38} a cubic vertex $a_0$ 
in the form
\begin{align}
\alpha_{_{{3}/{2}}} a_0^{susy} + y\,a_0^{cc}  &= \alpha_{_{{3}/{2}}}\,\Big(\bar{\psi}_\mu^\Delta \gamma^{\mu\nu\rho}  \sigma^{\alpha\beta} \nabla_{[\alpha} h_{\beta]\nu} \psi_\rho^\Omega k_{\Delta\Omega}\notag\\
&~~~~~~~~~ - h \bar{\psi}_\mu^\Delta \gamma^{\mu\nu\rho} \nabla_\nu \psi_\rho^\Omega k_{\Delta\Omega} + h_{\alpha\beta} \bar{\psi}^{\alpha\Delta} \gamma^{\beta\nu\rho} \nabla_\nu \psi_\rho^\Omega k_{\Delta\Omega} + h_{\alpha\beta} \bar{\psi}_\mu^\Delta \gamma^{\mu\alpha\nu} \nabla^\beta \psi_\rho^\Omega k_{\Delta\Omega}\notag\\
&~~~~~~~~~+ h_{\alpha\beta} \bar{\psi}_\mu^\Delta \gamma^{\mu\nu\alpha} \nabla_\nu \psi^{\beta\Omega} k_{\Delta\Omega} + \lambda h \bar{\psi}_\mu^\Delta \gamma^{\mu\nu} \psi_\nu^\Omega k_{\Delta\Omega} - 2 \lambda h_{\alpha\beta} \bar{\psi}^{\alpha\Delta} \gamma^{\beta\nu} \psi_\nu^\Omega k_{\Delta\Omega}\Big)\nonumber \\
& ~~~+ y \,\Big(\bar{\psi}^{\Delta}_{\mu} \left( F_a^{\mu\nu} - i\gamma^5 (*F_a)^{\mu\nu} \right) \psi^\Omega_{\nu} t^a_{\phantom{a}\Delta\Omega} - \lambda \bar{\psi}_{\Delta\rho} \gamma^{\mu\nu\rho} \psi_{\Omega\nu} A^a_\mu t_a^{\phantom{a}\Delta\Omega}\Big)\;.
\label{eq:3.48}
\end{align}

\paragraph{Yang-Mills algebra.}

The infinitesimal deformation corresponding to the Yang--Mills interactions, 
as obtained from the cohomological procedure \cite{Barnich:1993vg},
is well-known \cite{Barnich:1993pa}. We recall it here for completeness: 
\begin{equation}
a_2^{_{YM}} = \tfrac{1}{2} \,\mathcal{C}^\star_c \,f^c{}_{ab}\, \mathcal{C}^a \mathcal{C}^b \;,
\end{equation}
where $f^c{}_{ab} = -f^c{}_{ba}\,$. 
After solving the first order descent equations \eqref{eq:3.38}--\eqref{eq:3.40}, 
one obtains
\begin{align}
a_1^{_{YM}} &= A^{\star\mu}_{\phantom{\star}c} A_\mu^a \mathcal{C}^b f^c{}_{ab}\;,\\
a_0^{_{YM}} &= -\tfrac{1}{2} F_c^{\mu\nu} A_\mu^a A_\nu^b f^c{}_{ab}\;,
\end{align}
provided that
\begin{equation}
\label{cstrt4}
f_{abc} \coloneqq f^d{}_{bc} \delta_{ad} = f_{[abc]}\;.
\end{equation}

\paragraph{Total first order deformation of the gauge algebra.}

Summarising the results obtained so far, 
the total infinitesimal deformation of the gauge algebra that can be lifted 
up to a cubic vertex $a_0$ is given by
\begin{equation}
a_2 = \kappa\, a_2^{_{EH}} + \alpha_{_{{3}/{2}}} \,a_2^{susy} + y\, a_2^{cc-\Lambda} 
+ g_{_{YM}}\, a_2^{_{YM}}\;.
\label{eq:summarya2}
\end{equation}
It depends on the four independent deformation parameters 
$\kappa$, $\alpha_{_{{3}/{2}}}$, $y$ and $g_{_{YM}}\,$. 
We will see in the rest of this section how the 
existence of a second-order deformation will relate these four infinitesimal 
constants.
\vspace{.3cm}

Before turning to the second-order deformations, we want to classify 
the first-order deformations of the gauge transformations that do not 
modify the gauge algebra. The corresponding descent equations are
\begin{align}
\delta \hat{a}_1 + \gamma a_0 &= \sqrt{-\bar{g}} \;\nabla_\mu j^\mu_0\;,
\nonumber \\
\gamma \hat{a}_1 &=0\;.
\label{eq:abarun}
\end{align}
We present the classification of the possible terms in the following 
paragraphs.

\paragraph{Coupling of the vector fields to gravity.}

The following cocycle of $H(\gamma)$ at antifield number 2, 
\begin{equation}
\hat{a}_1^{_{G1}} = - k^{ab} \, A^{\star\mu}_{\phantom{\star}a} 
F_{b\mu\nu}\, \xi^\nu \;,
\end{equation}
can be lifted to the cubic vertex
\begin{equation}
{a}_0^{_{G1}} = -\tfrac{1}{8}\, h F_a^{\mu\nu} F_{b\mu\nu}\, k^{ab} 
-\tfrac{1}{2} \,F_a^{\mu\sigma} F_{b\sigma}^{\phantom{b\sigma}\nu} h_{\mu\nu}\,k^{ab}\;, 
\qquad k_{ab} = k_{ba}\;.
\end{equation}

\paragraph{Coupling of the spin-$\frac{1}{2}$ field to gravity.}

The cocycle 
\begin{equation}
\hat{a}_1^{_{G1/2}} = \chi^{\star}\, \gamma^{\mu\nu} \,\chi\, \nabla_{[\mu} \xi_{\nu]} 
+ 4 \,\chi^{\star}\, \nabla_\mu \chi \,\xi^\mu
\end{equation}
can be lifted through \eqref{eq:abarun} to yield the cubic vertex
\begin{equation}
a_0^{_{G1/2}} =  \bar{\chi}\, \gamma^\mu \nabla^\nu \chi \,h_{\mu\nu} 
- \bar{\chi}\, \gamma^\mu \nabla_\mu \chi\, h\;.
\end{equation}

\paragraph{Interaction between the spin-$\tfrac{3}{2}$, the spin-$1$ 
and the spin-$\mathbf{\frac{1}{2}}$ fields.}

Finally, the following cocycle of $H(\gamma)\,$, 
\begin{equation}
\hat{a}_1^{int} = \tfrac{1}{2}\, k_{a\Delta}\, \chi^{\star} F_{\mu\nu}^a \gamma^{\mu\nu} \zeta^\Delta + k^{a\Delta}\, A^{\star\mu}_{\phantom{\star}a} \bar{\zeta}_\Delta 
\gamma_\mu \chi\;,
\end{equation}
can be lifted in \eqref{eq:abarun} to produce the cubic 
interaction\footnote{\label{foonotemass}We could have 
allowed a nonzero mass term for the Majorana 
spin-$\tfrac{1}{2}$ field, since anyway it has no gauge invariance. 
In the case of a non-zero mass, however, it is easy to see 
that there is an obstruction to the existence of $a_0^{int}\,$. 
Since, on the other hand, second-order considerations show 
that this vertex is necessary for the consistency of a theory
that requires a Majorana spin-$\tfrac{1}{2}$ field, 
we find that the mass 
of the Majorana spin-$\tfrac{1}{2}$ field must vanish.}
\begin{equation}
\label{a0intspins}
a_0^{int} = \tfrac{1}{2} \,k_{a\Delta}\, \bar{\chi} \,
\gamma^\lambda \gamma^{\mu\nu} \,F_{\mu\nu}^a\, 
\psi_{\lambda}^\Delta\;.
\end{equation}

\paragraph{Total first order deformation of the gauge transformations.}
To summarise, the first-order infinitesimal deformations of the 
free theory \eqref{BVfunctionalAdS}, in the sector $a_1$ corresponding 
to the deformations of the gauge transformations, is given by the 
linear combination
\begin{equation}
a_1 = \kappa \,a_1^{_{EH}} + \alpha_{_{{3}/{2}}} \,
(a_1^{susy} +\tilde{a}_1^{_{G3/2}}) 
+ y \,(a_1^{cc-\Lambda} + \tilde{a}_1^{cc}) 
+ g_{_{YM}}\, a_1^{_{YM}} + \alpha_1 \, \hat{a}_1^{_{G1}} + \alpha_{_{1/2}}\, \hat{a}_1^{_{G1/2}} + \omega \, \hat{a}_1^{int}\;,
\label{eq:summarya1}
\end{equation}
which depends on seven independent deformation parameters:  
$\kappa\;,  \alpha_{_{3/2}}\;,  y \;, g_{_{YM}} \;,\alpha_1 \;, \alpha_{_{1/2}}$ and $\omega\,$.

\bigbreak

\subsection{General and specific quadratic constraints}
\label{subsection:3.4}

The goal of this subsection is to solve the 
second order master equation
\begin{equation}
s W_2 = -\frac{1}{2} \BV{W_1}{W_1}\;.
\label{eq:3.62}
\end{equation}
We expand $W_2$ in antifield number, 
$W_2 = \int d^4x\, \sqrt{-\bar{g}}\,(b_0 + b_1 + b_2)\,$, and insert 
the expression in the above equation that  
produces the following descent of equations:
\begin{align}
\label{db1ga0a1a0lambda}
\delta b_1 + \gamma b_0 &= - \sqrt{-\bar{g}}\,({a_1},{a_0}) 
+ \partial_\mu t^\mu_0\;,\\
\label{sgammaa1lambda}
\delta b_2 + \gamma b_1 &= -\frac{1}{2}\;\sqrt{-\bar{g}}\, ({a_1},{a_1}) 
-\sqrt{-\bar{g}}\, ({a_2},{a_1}) + \partial_\mu t^\mu_1\;,\\
\label{a2a2lambda}
\gamma b_2 &= - \frac{1}{2}\;\sqrt{-\bar{g}}\, ({a_2},{a_2}) + \partial_\mu t^\mu_2\;.
\end{align}

\paragraph{Descent equation of maximal antighost number.}

This paragraph is devoted to the resolution of equation 
\eqref{a2a2lambda}.
The latter resolution is straightforward, although rather lengthy, 
therefore we only list the obstructions encountered in the 
resolution of \eqref{a2a2lambda}.
The obstructions are elements of $H(\gamma)$ that remain non-trivial 
in $H(\gamma\vert d)\,$. 
In order to remove all the obstructions we have to impose some constraints 
between the deformation parameters and some constraints on the 
normalized coefficients. In the computation of \eqref{a2a2lambda}, 
other obstructions appear with independent structures, but multiplied by 
the same coefficients as those appearing below in 
\eqref{kalphaobstruction}--\eqref{YMobstruction}.
Therefore, setting these coefficients to zero kills several different 
obstructions simultaneously. 
The minimal set of obstructions that bring the relevant 
coefficients to be set to zero is:
\begin{align}
\label{kalphaobstruction}
{\cal O}_1 &= \alpha_{_{3/2}}\, \left( \alpha_{_{3/2}}\, k^{\Delta}{}_{\Sigma} k^{\Sigma}_{\phantom{\Sigma}\Omega} - \frac{\kappa}{4}\; k^{\Delta}{}_{\Omega} \right)\; \xi^{\star\mu} \nabla_{[\mu} \xi_{\nu]} \bar{\zeta}_\Delta \gamma^\nu \zeta^\Omega,\\
\label{nmkappaobstruction}
{\cal O}_2 &= \lambda \left( \alpha^2_{_{3/2}}\,2 k_{\Delta\Omega} k_{\Lambda\Gamma} 
- 2 y^2\, t_{a\Delta\Gamma} t^{a}_{\phantom{a}\Omega\Lambda} \right)\; (\zeta^{\star\Delta} \zeta^\Gamma)(\bar{\zeta}^{[\Omega} \zeta^{\Lambda]})\;,\\
\label{YMnmobstruction}
{\cal O}_3 &= \lambda y \left( 2 \lambda y ~ 2(t_{a})^\Delta_{\phantom{\Delta}\Sigma} (t_b)^\Sigma_{\phantom{\Sigma}\Omega} - g_{_{YM}} f^{c}{}_{ab} (t_c)^\Delta_{\phantom{\Delta}\Omega} \right)\; \zeta^\star_\Delta \zeta^\Omega C^{[a} C^{b]}\;,\\
\label{YMobstruction}
{\cal O}_4 &= - g_{_{YM}}^2 \,\left(f^c{}_{ab}\, f^a{}_{de}\right) \; C^\star_c C^{[b} C^d C^{e]}\;.
\end{align}
For these obstructions to vanish, a necessary condition we have to impose are 
the following relations on the deformation parameters: 
\begin{align}
\label{cstrt1}
\alpha_{_{3/2}} &= \frac{\kappa}{4}\;,\\
\label{cstrt2}
y^2 &= \frac{\kappa^2}{32},\\
\label{cstrt3}
g_{_{YM}} &= 2 \lambda y\;.
\end{align}
There thus remains only one free deformation parameter in the expression 
of $a_2\,$, in \eqref{eq:summarya2}, leaving us with four deformation parameters 
in total, taking into account those appearing in \eqref{eq:summarya1}.  
Then, in order for the four obstructions above to be identically zero we also 
have to impose constraints on the normalized coefficients 
involved in the definition of the $a_2$'s. 
First of all we must have 
\begin{equation}
\label{eq:3.73}
k^\Delta{}_{\Sigma} k^{\Sigma}_{\phantom{\Sigma}\Omega} = k^\Delta{}_{\Omega}\;.
\end{equation}
If $k^\Delta{}_{\Omega}$ is of maximal rank, that is to say 
if we choose that all the gravitini are coupled to gravity, 
then this normalised matrix is invertible 
and the only solution to the above constraint is
$k^\Delta{}_{\Omega} = \delta^\Delta_{\Omega}\,$. 
However, we will also consider the case 
where the matrix with components $k^\Delta{}_{\Omega}$
has the zero eigenvalue, with some multiplicity.
Another constraint that appears from the vanishing of the obstruction
\eqref{YMobstruction} is the Jacobi identity
\begin{equation}
\label{cstrt5}
f^c{}_{a[b}\, f^a{}_{de]}= 0\;,
\end{equation}
which tells us that $f^c{}_{ab}$ are the structure constants of a Lie algebra
for which $\delta_{ab}$ is an invariant tensor. From the complete antisymmetry and 
reality of $f_{abc}\,$, one recovers \cite{Barnich:1993pa} that the real Lie algebra 
must be semi-simple and compact.

Then, if we see the coefficients $t_a{}^{\Delta}{}_{\Omega}$ as $n_v$ 
square ${\cal N}_4\times {\cal N}_4\,$ matrices, we have the constraint
\begin{equation}
\label{cstrt6}
\left[ t_a , t_b \right]^{\Delta}{}_{\Omega} = f^c{}_{ab}\, (t_c)^{\Delta}{}_{\Omega}
\end{equation}
that means that these matrices characterise a representation of dimension $\mathcal{N}$ 
of the compact Lie group defined by the structure constants $f^c{}_{ab}\,$.
Finally, we have the completeness relation
\begin{equation}
\label{cstrt7}
(t^a)_{\Delta\Omega} (t_a)^{\Gamma\Sigma} = 
2 {k}^{[\Gamma}{}_{\Delta} {k}^{\Sigma]}{}_{\Omega}
\end{equation}
that also comes as a constraint in order to remove the obstruction \eqref{nmkappaobstruction} that appear in solving 
\eqref{a2a2lambda}. 
Once the above constraints are satisfied, we can solve \eqref{a2a2lambda} to
obtain the deformation of the gauge algebra, to second order in deformation:
\begin{align}
\label{eq:b2}
b_2 
&= \kappa^2 \, b_2^{_{EH}} + \frac{\kappa^2}{4} k^{\Delta\Omega} \;\zeta^{\star}_{\Delta} \gamma_{\mu\nu} \zeta_\Omega \left( \tfrac{1}{2}\,h^\mu_{\phantom{\mu}\sigma} \nabla^{[\nu} \xi^{\sigma]} + \xi_\sigma \nabla^{[\sigma} h^{\nu]\mu} \right) 
\notag\\ &~~
+\frac{\kappa^2}{8} k^{\Delta\Omega} \;\xi^\star_\mu \xi_\nu \bar{\zeta}_\Delta \gamma^{[\mu} \psi^{\nu]}_{\Omega} - \frac{\kappa^2}{32} k^{\Delta\Omega} k_{\Gamma\Sigma}\;\left( \zeta^{\star}_{\Delta} \gamma^{\mu\nu} \zeta_\Omega \right) \left( \bar{\zeta}^\Gamma \gamma_\nu \psi_{\mu}^\Sigma \right)\;, 
\end{align}
up to trivial terms and up to solutions $\tilde{b}_2$ to the homogeneous equation $\gamma \tilde{b}_2 = 0\,$. In the above expression, 
$b_2^{_{EH}}$ is the second-order deformation of the diffeomorphism algebra:
\begin{align}
    b_2^{_{EH}} = \tfrac{1}{4}\,\xi^{\star\alpha}\, \xi_\mu
    \Big[ h_{\alpha\nu} \nabla^{[\mu}\xi^{\nu]} - h^{\mu\nu} \nabla_{[\alpha}\xi_{\nu]} 
    - \nabla^{[\mu}h^{\nu]}{}_\alpha\, \xi_{\nu} \Big]\;,
\end{align}

\vspace{.3cm}

At this stage we can make two remarks on our preliminary results obtained in $AdS_4$ background:
\begin{itemize}
    \item[(i)] We can obtain the expression of the charge of the gravitini under the 
    gauge group in terms of the Newton constant and the cosmological constant by reading 
    off the coefficient in front of the minimal coupling terms in \eqref{eq:3.48}. 
    Using \eqref{cstrt2}, it is given by 
\begin{equation}
- \lambda y = \mp \frac{\lambda \kappa}{4\sqrt{2}}\;,
\end{equation}
where the ambiguity of sign translates the fact that the charge of the 
gravitini can have both signs;
\item[(ii)] Had we calculated the deformations around Minkowski spacetime, 
we would have found the minimal coupling of the gravitini 
to the gauge group at second order in deformation after  
the introduction of the first-order deformations 
\eqref{3.35}-\eqref{3.37}, 
since around flat space the cosmological constant is 
an infinitesimal deformation parameter.
Moreover, neither \eqref{nmkappaobstruction} nor \eqref{YMnmobstruction} would 
have appeared as obstructions during the resolution of the equation \eqref{a2a2}
around flat spacetime, since the latter expressions are multiplied by $\lambda\,$. 
Nevertheless we also find the constraint \eqref{cstrt2}, 
this time during the resolution of \eqref{sgammaa1}. 
Actually, perturbatively, we can recover all the 
deformations that one can obtain around Minkowski spacetime by taking the limit 
$\lambda \longrightarrow 0$ of the deformation around $AdS_4$ spacetime;
\item[(iii)] Independently of the matter content, the constraints 
\eqref{eq:3.73}--\eqref{cstrt7} must hold for the existence of any consistent theory involving spin-1 and spin-3/2 gauge fields.
\end{itemize}

\paragraph{Descent equation at antifield number 1.}

After having solved the equation \eqref{a2a2lambda} at antifield number 2, 
with the expression of $b_2$ given above, we can plug the result into 
the equation \eqref{sgammaa1lambda} and solve for $b_1\,$, the deformation 
of the gauge algebra at second order in deformation. If obstructions arise, 
the vanishing of the coefficients of the obstructions will give extra 
quadratic constraints on the deformation parameters and structure constants, 
in particular on the constants $\alpha_1\,$ and  
$\alpha_{_{1/2}}$ that appear in \eqref{sgammaa1lambda}. 
The ambiguity $\hat{b}_2$ in the solution \eqref{eq:b2} will affect 
$b_1$ through on-shell closure terms. These terms are expected to arise 
as the supersymmetry gauge algebra only closes on-shell, when no 
auxiliary fields are introduced.
On top of $\hat{b}_2$ terms with the general structure ``$\psi^\star\,\psi^\star\,\zeta\,\zeta\,$'' we also find that terms of the schematic 
form ``$\chi^\star A^\star \,\zeta\,\xi\,$'', for example, must be introduced in order to find the $b_1$ terms, namely, 
the deformations of the gauge transformations at second order in the 
fields.  

\subparagraph{Determination of $\boldsymbol{\alpha_1}$.}

We fix the value of the constant parameter $\alpha_1$ by inspection of 
the terms in \eqref{sgammaa1lambda} having the schematic form 
``$A^\star F \xi \xi\,$'', 
belonging to the  cohomological group $H(\gamma)\,$. 
These terms cannot come from $\delta b_2$ (because there is no $\mathcal{C}^\star$ in $b_2$) but can only come from $\BV{\hat{a}_1^{_{G1}}}{\hat{a}_1^{_{G1}}}$ and $\BV{a_2^{_{EH}}}{\hat{a}_1^{_{G1}}}\,$. 
Then, 
provided one adds the homogeneous piece 
$(\alpha_1)^2\; \tilde{b}_2^{_{G1}} = -\frac{(\alpha_1)^2}{2}\, \mathcal{C}^\star_a \xi^\mu \xi^\nu 
F^b_{\mu\nu} k^a_{\phantom{a}c} k^c_{\phantom{c}b}$ to $b_2\,$, 
the only obstruction that comes in the calculation of 
$-\frac{(\alpha_1)^2}{\sqrt{-\bar{g}}} \delta\tilde{b}_2^{_{G1}} -\frac{1}{2} \BV{\hat{a}^{_{G1}}_1}{\hat{a}^{_{G1}}_1} - \BV{a_2^{_{EH}}}{\hat{a}^{_{G1}}_1} + \nabla_\mu t^\mu_1$ is
\begin{equation}
\label{alpha1kappa}
{\cal O}_5 = \alpha_1 \left( \kappa \,k^a_{\phantom{a}b} - \alpha_1\, k^a_{\phantom{a}c} k^c_{\phantom{c}b} \right) A^{\star\sigma}_{\phantom{\star}a} \xi^\nu \nabla_{[\mu} \xi_{\nu]} F^{b\phantom{\sigma}\mu}_{\phantom{b}\sigma}\;.
\end{equation}
leading to the two constraints
\begin{align}
\label{cstrt8}
\alpha_1 = \kappa\;,\qquad 
k^a_{\phantom{a}b} = k^a_{\phantom{a}c}\, k^c_{\phantom{c}b}\;.
\end{align}
Provided the matrix with coefficients $k^a{}_b$ has maximal rank, 
meaning that all the spin-1 gauge field couple to gravity, 
the second constraint is solved by
$k^a_{\phantom{a}b} = \delta_a^b\,$, but otherwise, some vector 
fields can remain decoupled from gravity.

\subparagraph{Determination of $\boldsymbol{\alpha_{_{1/2}}}$.}

The determination of $\alpha_{_{1/2}}$ is realized by an analysis of the 
terms ``$\chi^\star \chi \nabla \xi \xi\,$'' that appear in the 
resolution of \eqref{sgammaa1lambda}. 
These terms can only arise from $\BV{\hat{a}_1^{_{G1/2}}}{\hat{a}_1^{_{G1/2}}}$ 
and $\BV{a_2^{_{EH}}}{\hat{a}_1^{_{G1/2}}}\,$. 
Here, three obstruction appears:
\begin{align}
{\cal O}_6 &= 4\alpha_{_{1/2}} \left( 4\alpha_{_{1/2}} - \kappa \right) \chi^\star \nabla_\mu \chi \xi_\nu \nabla^{[\mu} \xi^{\nu]}\;,\\
{\cal O}_7 &= \alpha_{_{1/2}}\left( 4\alpha_{_{1/2}} - \kappa \right)\,
\chi^\star \gamma^{\mu\alpha}\chi \bar{g}_{\beta\nu}\,\nabla_{[\mu}\xi_{\nu]}
\nabla_{[\alpha}\xi_{\beta]}\;,\\
{\cal O}_8 &= -\alpha_{_{1/2}}\lambda^2\left( 4\alpha_{_{1/2}} - \kappa \right)\, \chi^\star \gamma^{\mu\nu}\chi \xi_\mu \xi_\nu\;.
\end{align}
The coefficients of these obstructions all vanishes if we set
\begin{equation}
\label{cstrt9}
\alpha_{_{1/2}} = \frac{\kappa}{4}\;.
\end{equation}

\subparagraph{Determination of $\boldsymbol{\omega}$.}

In order to determine $\omega\, k_{a\Delta}$ we calculate all the terms that 
have the general structure ``$A^\star F \zeta \zeta\,$''. 
They only come from $\BV{\hat{a}_1^{int}}{\hat{a}_1^{int}}\,$, $\BV{a_1^{cc-\Lambda}}{\tilde{a}_1^{cc}}$ and $\BV{a_2^{susy}}{\hat{a}_1^{_{G1}}}\,$. 
Focusing on the ``$A^\star F \zeta \zeta\,$'' terms, we have 
\begin{align}
-\frac{1}{2} \BV{\omega \,\hat{a}_1^{int}}{\omega \,\hat{a}_1^{int}}|_{A^\star F \zeta \zeta} &= 
- \frac{\omega^2}{2} A^{\star a}_{\phantom{\star}\mu} F_{\rho\sigma}^b \bar{\zeta}^{\Delta} \gamma^\mu \gamma^{\rho\sigma} \zeta^\Omega k_{a\Delta} k_{b\Omega}\;,
\\
-\frac{1}{2} \BV{y\, a_1^{cc-\Lambda}}{y\,\tilde{a}_1^{cc}}|_{A^\star F \zeta \zeta} &= 
\frac{\kappa^2}{32}\, A^{\star a}_{\phantom{\star}\mu} F_{\rho\sigma}^b \bar{\zeta}^{\Delta} \gamma^{\rho\sigma} \gamma^\mu \zeta^\Omega t_{a\Sigma\Delta} t_{b\phantom{\Sigma}\Omega}^{\phantom{a}\Sigma}\;,
\\
-\BV{\alpha_{_{3/2}} \,a_2^{susy}}{\alpha_1 \,\hat{a}_1^{_{G1}}} &= 
-\frac{\kappa^2}{16} A^{\star\mu}_{\phantom{\star}a} F^a_{\mu\nu} \bar{\zeta}_\Delta \gamma^\nu \zeta^\Delta\;.
\end{align}
By using the identities \eqref{gamma2gamma1gamma3gamma1} and \eqref{gamma1gamma2gamma3gamma1}, the RHS of
the following equation 
\begin{equation}
\tfrac{1}{\sqrt{-\bar{g}}} \left( \gamma b_1 \right)|_{A^\star F \zeta \zeta} = -\tfrac{1}{\sqrt{-\bar{g}}} \left( \delta b_2 \right)|_{A^\star F \zeta \zeta} - \frac{1}{2} \BV{a_1}{a_1}|_{A^\star F \zeta \zeta} - \BV{a_2}{a_1}|_{A^\star F \zeta \zeta} + \left( \nabla_\mu t^\mu_1 \right)|_{A^\star F \zeta \zeta}
\end{equation}
is
\begin{align}
A^{\star\mu}_{\phantom{\star}a} F_{b\mu\nu} \bar{\zeta}_{(\Delta} \gamma^\nu \zeta_{\Omega)} \left( \frac{\kappa^2}{16} k^{ab} k^{\Delta\Omega} - \frac{\kappa^2}{16} (t^a)^{\Sigma(\Delta} (t^{|b|})_\Sigma^{\phantom{\Sigma}\Omega)} - \omega^2 k^{a(\Delta} k^{|b|\Omega)} \right) \notag\\
\label{AstarFzetazeta}
+ A^{\star\mu}_{\phantom{\star}a} F_{b}^{\rho\sigma} \bar{\zeta}_{[\Delta} \gamma_{|\mu\rho\sigma|} \zeta_{\Omega]} \left( \frac{\kappa^2}{32} (t^a)^{\Sigma[\Delta} (t^{|b|})_\Sigma^{\phantom{\Sigma}\Omega]} - \frac{\omega^2}{2} k^{a[\Delta} k^{|b|\Omega]} \right) + \nabla_\mu j_1^\mu\;.
\end{align}
They give obstructions to finding $b_1$ because it is in $H(\gamma |d)\,$. 
Therefore the coefficient in front of these obstructions must vanish identically,  
providing us with the following constraints on the deformation structures:
\begin{align}
\label{cstrt10}
\omega^2 &= \frac{\kappa^2}{16}\;,\\
\label{cstrt11}
k^{ab} k^{\Delta\Omega} &= (t^a)^{\Sigma(\Delta} (t^{|b|})_\Sigma^{\phantom{\Sigma}\Omega)} + k^{a(\Delta} k^{|b|\Omega)}\;,\\
\label{cstrt12}
k^{a[\Delta} k^{|b|\Omega]} &= (t^a)^{\Sigma[\Delta} (t^{|b|})_\Sigma^{\phantom{\Sigma}\Omega]}\;.
\end{align}

As regards the number of deformation parameters, after the resolution of the master equation at first order we had seven independent ones. The three constraints \eqref{cstrt1}, \eqref{cstrt2} and \eqref{cstrt3} coming from the resolution of the master equation at second order and antifield number 2 \eqref{a2a2} lower the number of deformation parameters down to four. Then the three constraints \eqref{cstrt8}, \eqref{cstrt9} and \eqref{cstrt10} coming in the resolution at antifield number 1 \eqref{sgammaa1} leave us with only one deformation parameter: $\kappa\,$.
Of course, one can always take the $\lambda\longrightarrow 0$ limit of the final 
result and also view the cosmological constant as a deformation parameter. 
This is certainly the correct point of view when performing the deformation around 
flat spacetime, but less natural when performing the deformation around the $AdS_4$
background, where the cosmological constant can be arbitrarily large in absolute 
value.

\paragraph{Quadratic constraints.}

Let us summarise all the quadratic constraints on the normalised coefficients arising during the resolution of the master equation at second order:

\begin{align}
\label{GeneralConstraints}
    k^\Delta{}_{\Sigma} k^{\Sigma}_{\phantom{\Sigma}\Omega} &= k^\Delta{}_{\Omega}\;,\quad 
    f^c{}_{a[b}\, f^a{}_{de]} = 0\;,\quad 
    2(t_{a})^\Delta_{\phantom{\Delta}\Sigma} (t_b)^\Sigma_{\phantom{\Sigma}\Omega} = f^{c}{}_{ab} (t_c)^\Delta_{\phantom{\Delta}\Omega}\;,\quad
    (t^a)_{\Delta\Omega} (t_a)^{\Gamma\Sigma} = 
2 {k}^{[\Gamma}{}_{\Delta} {k}^{\Sigma]}{}_{\Omega}\;,\\
\label{SpecificConstraints}
    k^a_{\phantom{a}c}\, k^c_{\phantom{c}b} &= k^a_{\phantom{a}b}\;,\quad
    k^{ab} k^{\Delta\Omega} = (t^a)^{\Sigma(\Delta} (t^{|b|})_\Sigma^{\phantom{\Sigma}\Omega)} + k^{a(\Delta} k^{|b|\Omega)}\;,\quad
k^{a[\Delta} k^{|b|\Omega]} = (t^a)^{\Sigma[\Delta} (t^{|b|})_\Sigma^{\phantom{\Sigma}\Omega]}\;.
\end{align}

The constraints \eqref{GeneralConstraints} are completely general in the sense that they 
are valid for arbitrary numbers of vectors and Rarita-Schwinger gauge fields, and 
independent of the matter content. The sole restriction is on the singleness of the 
graviton. On the other hand the constraints \eqref{SpecificConstraints} will 
be modified by considering more than a single Majorana spin-1/2 field and some 
scalar fields.
\vspace{.3cm}

Depending on the number of gravitini coupled to gravity, that is to say the rank of the 
matrix of coefficients $k^\Delta{}_\Sigma\,$, 
different situations can arise as solution of the full set of constraints 
\eqref{GeneralConstraints}--\eqref{SpecificConstraints}. 
In the following we detail the resolution of these constraints 
for a number of Rarita-Schwinger gauge fields coupled to 
gravity less than or equal to three: $\mathcal{N}_4\leqslant 3\,$. 
This will lead to the uniqueness of 
$\mathcal{N}_4 \leqslant 3$ pure supergravities, 
when the field content is rich enough. 
The theory that describes the coupling of the $\mathcal{N}_4=1$ vector 
supermultiplet (1,1/2) to the $\mathcal{N}_4=1$ supergravity 
multiplet (2,3/2) is also included in the solution of the constraints 
when the menu of fields 
is adapted. A richer matter content is required to complete the 
specific constraints \eqref{SpecificConstraints} in order to be able to include other matter-coupled 
supergravities and higher extended supergravity theories as solutions, 
a problem that we will treat elsewhere and that should reproduce, for 
instance, the models found in 
\cite{Cremmer:1978hn,Cremmer:1982en,Cremmer:1984hj}.

\subparagraph{Uniqueness of $\boldsymbol{\mathcal{N}_4 = 3}$ pure sugra.}

Firstly we will suppose that the matrix of coefficients $k^\Delta{}_\Sigma$ has 
rank 3. 
Then from \eqref{eq:3.73} we have $k^\Delta{}_\Sigma = \delta^\Delta_\Sigma$ 
when $\Delta$ and $\Sigma$ run over $\{1,2,3\}\,$. The coefficients 
$k^\Delta{}_\Sigma$ are all null otherwise. 
It thus means that three of the massless spin-$\tfrac{3}{2}$ fields 
are coupled to gravity and therefore are gravitini. 
In order to study the uniqueness of $\mathcal{N}_4 = 3$ supergravity, 
let us see how three of the vector gauge fields couple to them.
Concerning the Yang-Mills coupling of these three vector gauge fields, 
the two constraints \eqref{cstrt4} and \eqref{cstrt5} tell us that the 
coefficients $f^c{}_{ab}$ are the structure constants of a compact, 
real and semi-simple Lie group of dimension $n_v = 3\,$, namely $SO(3)\,$.
Then the two constraints \eqref{cstrt6} and \eqref{cstrt7} tell us 
that the matrices $(t_a)^{\Delta\Omega}$ form a representation of 
dimension $\mathcal{N}_4 = 3$ of $SO(3)$ and is a complete basis 
of the antisymmetric $\mathcal{N}_4\times\mathcal{N}_4 = 3\times 3$ 
matrices. All in all we have to take both $f^c{}_{ab}$ 
and $t_a^{\phantom{a}\Delta\Omega}$ to be the 3D Levi-Civita antisymmetric symbol
and from now on we substitute the Latin indices $i,j,\ldots$ for the Latin indices 
$a,b,\ldots$ and for the upper-case Greek indices:
\begin{equation}
(f_{ab}^{\phantom{ab}c} \;, t_a^{\phantom{a}\Delta\Omega}) \longrightarrow (\epsilon_{ij}^{\phantom{ij}k} \;, \epsilon_i^{\phantom{i}jk})\;.
\end{equation}
With these notations, 
the remaining two constraints \eqref{cstrt11} and \eqref{cstrt12} 
that we have to solve can be written as
\begin{align}
\delta^{ij} \delta^{kl} &= \epsilon^{ip(k} \epsilon^{|j|\phantom{p}l)}_{\phantom{|j|}p} + k^{i(k} k^{|j|l)}\;,\\
k^{i[k} k^{|j|l]} &= \epsilon^{ip[k} \epsilon^{|j|\phantom{p}l]}_{\phantom{|j|}p}\;.
\end{align}
By use of the identity on the Levi-Civita symbols and Kronecker delta's
\begin{equation}
\epsilon^{p_1 \dots p_k i_1\dots i_l} \epsilon_{p_1 \dots p_k j_1 \dots j_l} 
= k! \,l! \,\delta^{[i_1}_{\phantom{[}j_1} \dots \delta^{i_l]}_{j_l}\;,
\end{equation}
we can rewrite these two constraints in the following simple form
\begin{align}
\label{fcstrt1}
k^{i(k} k^{|j|l)} &= \delta^{i(k} \delta^{|j|l)}\;,\quad
k^{i[k} k^{|j|l]} = \delta^{i[k} \delta^{|j|l]}\;,
\end{align}
where $k^{ij}$ is now the notation for the coefficients $k^{a\Delta}$.
In this way, it is easy to see that all the solutions
to \eqref{fcstrt1} are given by $k^{ij} = \pm \delta^{ij}\,$.
The $\pm$ sign can be absorbed in the definition of $\omega$ in term 
of $\kappa\,$, cf. \eqref{cstrt10}. 
Finally, we choose
\begin{equation}
k^{ij} = \delta^{ij}\;,
\end{equation}
which is exactly what appears in the cubic vertex involving the spins $(\tfrac{1}{2},1,\tfrac{3}{2})$ in $\mathcal{N}_4 = 3$ pure supergravity \cite{Freedman:1976aw,Freedman:1976nf}.

\subparagraph{Uniqueness of $\boldsymbol{\mathcal{N}_4 = 2}$ pure sugra.}

Since the $\mathcal{N}_4 = 2$ supergravity theory (without cosmological constant) is 
already discussed in the subsection \ref{subsec:ConsistentFlat}, we will be very brief 
here. The solution of the constraints \eqref{GeneralConstraints} and 
\eqref{SpecificConstraints}, which hold in the case where the cosmological constant is 
negative, can be analysed in the case where the matrix of coefficients $k^\Delta{}_\Sigma$ 
is of rank 2, and has to be identified with $\delta^\Delta_\Sigma$ when the indices run 
over the two integer values $\{1,2\}\,$. 
The $\mathcal{N}_4 = 2$ supergravity theory \cite{Freedman:1976aw} is recovered by 
considering the coupling with a single vector gauge field and no Majorana spin-1/2. 
Indeed 
the only non-trivial coefficients apart from $k^\Delta{}_\Sigma$ are therefore 
$t_a{}^{\Delta\Omega}$ and one can easily see that all the constraints are satisfied 
if and only if it is identified with $\epsilon^{\Delta\Omega}$.

\subparagraph{$\boldsymbol{\mathcal{N}_4 = 1}$ sugras.}

Finally we will study the case where the matrix of coefficients $k^\Delta{}_\Sigma$ 
has rank 1. The solution of the constraint \eqref{eq:3.73} therefore is 
$k^{1}{}_{1}=1$ as sole non-zero component of $k^\Delta{}_\Sigma\,$. 
The simplest solution to all the constraints 
is to take all the other coefficients to vanish. This almost trivial solution is possible 
only when a single real Rarita-Schwinger gauge field 
couples to gravity because the RHS of \eqref{cstrt7} vanishes only in this case. 
This yields the $\mathcal{N}_4 = 1$ supergravity theory, 
with \cite{Townsend:1977qa} and without \cite{Freedman:1976xh,Deser:1976eh} 
cosmological constant.

Still in the case where there is a single gravitino, another  
solution can be found. From \eqref{cstrt7} we will still set the coefficients 
$t_a{}^{\Delta\Omega}$ to zero. 
Then the constraints \eqref{SpecificConstraints} become
\begin{equation}
    k^a_{\phantom{a}c}\, k^c_{\phantom{c}b} = k^a_{\phantom{a}b}\;,\quad
    k^{ab} k^{\Delta\Omega} = k^{a(\Delta} k^{|b|\Omega)}\;,\quad
k^{a[\Delta} k^{|b|\Omega]} = 0\;,
\end{equation}
that we solve by considering that a single vector gauge field 
couples to gravity by taking $k^a_{\phantom{a}b}=1\,$ for $a=b=1\,$, 
and $k^a_{\phantom{a}b}=0$ otherwise.  
Subsequently, the coefficients $k^{a\Delta}$ have to be 1 when $a=1$ and $\Delta=1\,$,
and 0 otherwise. This leads to a theory, with negative cosmological constant terms, 
of a vector supermultiplet (1,1/2) coupled to the supergravity multiplet (2,3/2) 
through the vertex \eqref{a0intspins}. 
The flat limit of this model is none other than the model of \cite{Ferrara:1976um}.

\paragraph{Summary and higher-order deformations.}

To summarise, we reproduced all the cubic vertices of, 
respectively, ${\mathcal{N}_4 = 3}$, 
${\mathcal{N}_4 = 2}$ and ${\mathcal{N}_4 = 1}$ 
pure supergravities, as well as the cubic vertices of the 
model of \cite{Ferrara:1976um} including extensions by cosmological 
constant terms that had not been discussed before in the literature, 
to the best of our knowledge. 
This was done by setting 
$W = W_0 +g\, W_1(\tilde{c}) + {\cal O}(g^2)$ 
for a set of infinitesimal parameters collectively denoted by 
$g\,$ and for some coupling constants collectively denoted 
by $\tilde{c}\,$, $W_0$ being the solution to the master equation 
$(W_0,W_0)=0$ for the free theory, 
and by classifying the general solution to the 
equation $(W,W)=0$ at first order in $g\,$, 
under the assumptions stated in our theorem.
We denote the general solution by $W_1^{sugra}(\tilde{c})\,$ 
as it contains, in its antifield-independent piece, all the cubic 
vertices of the supergravity theories with
${\cal N}_4\leqslant 3\,$ listed above.
It depends linearly on the parameters $\tilde{c}\,$ that 
must obey linear constraints, as for example $f_{abc}=f_{[abc]}$
for the Yang-Mills deformation. 

Having classified the solution space for $W_1\,$, we then 
turned to the master equation at second order in $g\,$, 
$(W_0,W_2)+\tfrac{1}{2}\,(W_1,W_1)=0\,$, and solved it for $W_2$
at antifield numbers 2 and 1. This enabled us to find quadratic 
constraints on the parameters $\tilde{c}\,$, allowing 
four classes of nontrivial solutions, depending on the rank of the 
matrices $k^\Delta{}_\Gamma\,$ and $k^a{}_b\,$ that are part of 
the constants denoted by $\tilde{c}\,$.
The solution $W_2$ at antifield number zero will contain the usual 
quartic vertices of the various supergravity theories discussed 
above, although we have not computed them explicitly, as this is 
not necessary to prove the uniqueness, under the assumptions 
of our theorem, of the aforementioned supergravity theories 
with ${\cal N}_4\leqslant 3\,$. 

Indeed, knowing $W_1^{sugra}(\tilde{c})\,$, 
the general solution of the order-$g$ equation 
$(W_0,W_1)=0$ --- under the assumptions stated in our theorem --- 
and knowing the full theories that precisely reproduce these 
first-order interactions, one can deduce that the general solution 
of the master equation, to all orders, will just produce
the nonlinear theories whose cubic vertices are parts of our general 
solution $W_1$ at antifield number zero.
We can repeat, in our current context, the general argument of the 
section 7 of \cite{Boulanger:2000rq} leading to the uniqueness 
of Einstein-Hilbert's action from the uniqueness of its cubic vertex. 
From the results obtained so far, we have that 
$W = W_0 + g\,W_1^{sugra}(\tilde{c}) + g^2\,W_2+\ldots\,$,
where the various parameters, collectively denoted by 
$\tilde{c}\,$, solve the quadratic constraints
\eqref{GeneralConstraints} and \eqref{SpecificConstraints}.
Therefore, depending on the class of solution of these quadratic 
constraints, $W_1^{sugra}(\tilde{c})$ denotes the solution of 
the master equation at first order in interactions, for one of the 
four ${\cal N}_4\leqslant 3$ supergravity theories recalled 
above.  

The functional $W_2$ has to solve $sW_2 = -\tfrac{1}{2}\,(W_1^{sugra}(\tilde{c}),W_1^{sugra}(\tilde{c}))\,$, 
therefore one can set\footnote{The notation  $W_2^{sugra}(\tilde{c}^2)$ 
is meant to indicate 
that $W_2^{sugra}$ depends quadratically on the parameters.} 
$W_2 = W_2^{sugra}(\tilde{c}^2)+W'_2\,$ , 
for $W_2^{sugra}$ the second-order solution of the classical master 
equation
for the corresponding nonlinear ${\cal N}_4\leqslant 3\,$ 
supergravity theory that we know exists.
The functional $W_2'$ must solve the homogeneous equation $(W_0,W_2')=0\,$, whose general solution (under the assumptions 
stated in our theorem) is $W_2' = W^{sugra}_1({c}')$ 
featuring some parameters ${c}'\,$ that solve the same linear 
constraints as the parameters $\tilde{c}\,$. 
The equation for $W_3$ is then 
$(W_0,W_3) = - (W_2 , W_1^{sugra}(\tilde{c}))$
$=-(W_2^{sugra}(\tilde{c}^2),W_1^{sugra}(\tilde{c}))$
$-(W_1^{sugra}(\tilde{c}),W_1^{sugra}({c}'))\,$.
We can set $W_3 = W_3^{sugra}(\tilde{c}^3) + W_3'$
where $W_3'$ must solve 
\begin{equation}
(W_0 , W_3') = - (W_1^{sugra}(\tilde{c}),W_1^{sugra}({c}'))\;.  
\label{eqHOrder}
\end{equation}
The right-hand side of the latter equation can be $s\,$-exact 
only if the parameters $\bar{c}\coloneqq\tilde{c}+g\,{c}'$ satisfy 
the quadratic constraints \eqref{GeneralConstraints} and 
\eqref{SpecificConstraints}, to the relevant order in $g\,$. 
From $W_2' = W_1^{sugra}({c}')\,$, eq. \eqref{eqHOrder} 
and eq. \eqref{eq:3.62}, one finds that 
$W_3' = 2 \,W_2^{sugra}(\tilde{c}{c}')$ 
up to a solution $W_3{}''$ of the equation 
$(W_0,W_3{}'') = 0\,$, i.e. $W_3{}'' = W_1^{sugra}({c''})\,$. 
Putting things together and continuing to higher orders, 
one obtains
$W^{sugra}= W_0 + W_1^{sugra}(c) + W_2^{sugra}(c^2) + \ldots$
that solves $(W^{sugra},W^{sugra})=0$ and depends on 
the coupling constants 
$c = g\,\tilde{c} + g^2\,c'+ g^3\,c''+ \ldots\,$.
This proves the uniqueness of the corresponding four 
supergravity theories discussed above.  

Therefore, knowing the existence of the 
complete supergravity theories whose cubic vertices we classified 
and identified above is enough to deduce the uniqueness of the 
corresponding nonlinear theories. Still, it is interesting to exhibit 
the quartic vertices from the cohomological approach. We have already 
provided quartic terms in the structure of the gauge algebra, including the typical on-shell closure terms that are a landmark of all 
supergravity 
theories without auxiliary fields. We intend to give them all, as well 
as the typical quartic vertices, in a forthcoming publication, although
we stress that this is not at all necessary for the proof of the uniqueness
of these models.

\section{Conclusions and perspectives}
\label{sec:Outlooks}

In this note we used the purely algebraic reformulation 
\cite{Barnich:1993vg} of the problem of introducing consistent interactions 
among (gauge) fields in order to study theories that couple 
a set of massless spin-$\tfrac{3}{2}$ fields to a set 
of Maxwell fields in four dimensions. This work can therefore 
be seen as the extension of the analyses performed in 
\cite{Boulanger:2000rq} and \cite{Boulanger:2001wq}. 
We recovered and proved the uniqueness of all the known, pure supergravity 
theories in four dimensions with $\mathcal{N}_4\leqslant 3\,$ 
and also obtained strong necessary constraints on the gaugings of 
higher matter-coupled or more extended supergravity models. 
In a future work, we will increase the menu of matter fields and 
study the solution of these constraints.
In perturbation around $AdS_4$ background for pure ${\cal N}_4\geqslant 4$
models, it will be interesting to see how both the possible gaugings and 
exponentials of the scalar fields appear in the appropriate cohomologies. 
An interesting perspective for future work would also be to 
increase the spacetime dimension and consider $p\,$-form gauge 
fields in the spectrum, 
so as to understand, from a BRST-cohomological point of view, 
the role of the latter fields in higher-dimensional, extended 
gauged supergravities.

\section*{Acknowledgements}
We thank Glenn Barnich, Marc Henneaux, Victor Lekeu and Arash 
Ranjbar for discussions.
Nicolas Boulanger thanks the Laboratoire de Physique Th\'eorique de l’\'Ecole Normale 
Sup\'erieure de Paris for kind hospitality, while Bernard Julia thanks the 
Groupe de M\'ecanique et Gravitation of UMONS for hospitality. 
This work was partially supported by FNRS-Belgium (convention PDR 
T.1025.14).

\appendix

\section{Conventions and some Diracology}
\label{Diracology}

In this appendix, we spell out our conventions for spinors in Minkowski space 
$\mathbb{R}^{1,3}$ and recall some known identities on Dirac matrices. 
A useful compendium of tools for supersymmetry, including conventions and 
identities on Dirac matrices in various dimensions, can be found in 
\cite{VanProeyen:1999ni}.

\paragraph{Clifford Algebra.} Our convention for the Levi-Civita 
tensor $\epsilon^{abcd}$ in Cartesian coordinates 
is such that $\epsilon^{0123}=1=-\epsilon_{0123}\,$. We use the mostly 
plus convention for the Minkowski metric in Cartesian coordinates: 
$\eta=$ diag $(-,+, + ,+)\,$. The Clifford algebra 
\begin{equation}
\label{Clifford}
\{\gamma^a,\gamma^b\}=2\eta^{ab}\Id_{4}
\end{equation}
reads $\{\gamma^a,\gamma^b\}_A^{\phantom{A}B}=2\eta^{ab}\delta_{A}^{B}\,$
in components, with $A,B=1,\dots, 4\,$ and where we take as a convention 
that the components of the gamma matrix $\gamma^a$
are $(\gamma^a)_A^{\phantom{A}B}\,$.  

\paragraph{Dirac adjoint.}

Our conventions for the complex, transpose and Hermitian 
conjugates of a matrix $M$ in terms of its components $M_{A}{}^{B}$ are  
$[M^*]{}^{A'}{}_{B'} = (M_A^{\phantom{A}B})^*\,$, 
$[M^T]{}^B{}_A= M_A{}^{B}$ and $[M^\dagger]{}_{B'}{}^{A'} = (M_A{}^{B})^*\,$. 
The components of a Dirac spinor are denoted by $\psi_A\,$, $A=1,\dots, 4\,$.
A sesquilinear form on the space of Dirac spinor is introduced,
$\beta_{A'}^{\phantom{A'}B} \coloneqq i (\gamma^0)_A^{\phantom{A}B}\,$, 
from which the Dirac adjoint of the Dirac spinor $\psi_A$ is defined, 
as usual, by
$\bar{\psi}^A \coloneqq (\psi^\dagger)^{B'} \beta_{B'}^{\phantom{B'}A}\,$
in components, or $\bar{\psi} = \psi^\dagger \beta\,$ in matrix notation. 

\paragraph{}

Different sets of matrices obeying the Clifford algebra \eqref{Clifford} 
are related by a change of basis. As the matrices $-(\gamma^a)^\dagger$ 
also satisfy the relation \eqref{Clifford}, they can be obtained 
from a change of basis starting from $\gamma^a\,$. 
Indeed, one has 
$\beta \gamma^a \beta^{-1} = -(\gamma^a)^\dagger\,$.
The matrix $\gamma^5$ is defined as
$\gamma^5 \coloneqq i \gamma^0 \gamma^1 \gamma^2 \gamma^3\,$.
It obeys $(\gamma^5)^2 = \Id_4$ and anticommutes with the four 
matrices of the Clifford algebra $\gamma^a\,$.
One also defines
$\gamma^{a_1 a_2 \dots a_n} \coloneqq \gamma^{[a_1} \gamma^{a_2} \dots \gamma^{a_n]}\,$, 
where 
$$[a_1\dots a_n] = \frac{1}{n!} \sum\limits_{\sigma \in S_n} (-1)^{Sgn(\sigma)} a_{\sigma(1)} \dots a_{\sigma(n)}$$ defines the complete antisymmmetrization with 
strength one.
We denote the matrices $\gamma^{a_1 a_2 \dots a_n}$ by $\gamma^{a[n]}\,$, 
for the sake of brevity. 
As usual, one gathers together these 16 matrices in the following 
set $\{ M_I \}_{I=1\dots 16} = \{ \Id_4, \gamma^a, \gamma^{a[2]}, \gamma^{a[3]}, \gamma^5 \}$ that forms a basis of the $4 \times 4$ complex matrices.

\paragraph{Majorana conjugate.}

Another representation of the Clifford algebra is given by minus 
the transposed Dirac matrices $-(\gamma^a)^T$ with components 
$-[(\gamma^a)^T]{}^{A}{}_{B}\,$. 
The charge conjugation matrix $C^{AB}$ realizes the similitude 
\begin{equation}
\label{CliffC}
C \gamma^a C^{-1} = -(\gamma^a)^T\;.
\end{equation}
In even dimension it is possible to choose $C$ to be unitary $C^\dagger=C^{-1}$ and 
antisymmetric $C^T = -C\,$.

Starting from the 16 basis matrices $M_I$ defined above, 
one deduces from \eqref{CliffC} and the antisymmetry of the charge conjugation 
matrix that $C \gamma^a$ and $C \gamma^{a[2]}$ are symmetric matrices, 
whereas $C \gamma^{a[3]}$, $C \gamma^5$ and $C$ are antisymmetric. 
The charge conjugation matrix $C$ is instrumental in providing 
a definition of the Majorana conjugate  
$\widetilde{\psi} \coloneqq \psi^{T} C\,$, 
or $\widetilde{\psi}^A = \psi_B C^{BA} = - C^{AB} \psi_B\,$ in components.
A Majorana spinor is a spinor such that its Dirac conjugate $\bar{\psi}^A$ 
is equal to its Majorana conjugate $\widetilde{\psi}^A\,$.

\paragraph{Identities relative to the Dirac matrices.}

We collect here various identities that we needed in our calculations:
\begin{align}
\label{gamma3gamma2}
\gamma^{mnr}\gamma^{ab} &\equiv \overset{(1)}{\left( \gamma^{mnr}\gamma^{ab} \right) }
+ \overset{(3)}{\left( \gamma^{mnr}\gamma^{ab} \right) }\;,
~~~{\rm where}\\
\nonumber
\overset{(1)}{\left( \gamma^{mnr}\gamma^{ab} \right) } 
\coloneqq -6 \eta^{a[m} \eta^{|b|n} &\gamma^{r]}~~{\rm and}~~
\overset{(3)}{\left( \gamma^{mnr}\gamma^{ab} \right) } \coloneqq 
3 \left( \eta^{a[m} \gamma^{nr]b} - \eta^{b[m} \gamma^{nr]a} \right)\;.
\end{align}
Moreover, 
\begin{align}
\label{gamma3gamma1gamma2}
\gamma_{mnp} \gamma^p &= 2 \gamma_{mn}\;,
\\
\label{gamma2gamma1gamma3gamma1}
\gamma_{np} \gamma_m &= \gamma_{npm} - 2\eta_{m[n} \gamma_{p]}\;,
\\
\label{gamma1gamma2gamma3gamma1}
\gamma_{n} \gamma_{ab} &= \gamma_{nab} + 2\eta_{n[a} \gamma_{b]}\;,
\\
\label{commgamma1gamma2gamma1}
[\gamma_{n} \,, \gamma_{ab}] &= 4\eta_{n[a} \gamma_{b]}\;,
\\
\label{gamma2gamma2}
\gamma_{ab}\gamma_{mn} &= -i \gamma^5 \epsilon_{abmn} 
- 2 (\eta_{m[a}\,\gamma_{b]n}-\eta_{n[a}\,\gamma_{b]m})
- 2 \eta_{m[a} \eta_{b]n}\Id\;.
\end{align}
Finally, 
\begin{align}
\label{gamma3epsgamma5}
\gamma_{mnp} &= i \epsilon_{mnpq} \gamma^{q}\gamma^5 = -i \epsilon_{mnpq} \gamma^5\gamma^{q}\;.
\end{align}

\section{Appearance of conserved currents in gaugings}
\label{Hmoins1}

In this Appendix we make some comments on the relation between what is 
known as the Noether procedure for introducing couplings among fields 
and the BRST-BV reformulation \cite{Barnich:1993vg,Henneaux:1997bm} 
of the consistent coupling procedure explained e.g. in \cite{Berends:1984rq}.
More precisely, in the present paper we have followed the 
techniques and used some general theorems proved in 
\cite{Barnich:1994db,Barnich:1994mt}. According to this procedure
and the equations \eqref{da1ga0}--\eqref{ga20} for the determination 
of the first-order interactions, 
one can classify all such infinitesimal 
interactions, represented in the Lagrangian by $a_0\,$, according to whether
(i) $a_2=0=a_1\,$, (ii) $a_2=0$ but $a_1\neq 0$ and (iii) $a_2\neq 0\,$. 
In all three cases, one can write the vertex $a_0$ in the Noether form 
``gauge field times conserved current''. 
\vspace{.3cm}

From the results we obtained in the body of the paper, we now discuss these 3 cases 
in a way that closely follows the exposition made in section 3.2 
of \cite{Barnich:2017nty}, but this time for a spectrum of fields 
that also contains spin-2, spin-$\tfrac{3}{2}$ and 
spin-$\tfrac{1}{2}$ fields. A detailed discussion of 
the pure spin-2 case was given in section 11 of~\cite{Boulanger:2000rq}.
\begin{itemize}
    \item[(i)] \underline{$a_2=0=a_1$.} In these cases the vertices 
and hence the currents can be written as functions of the (derivatives of the) 
field strengths of the various fields and of the spin-$\tfrac{1}{2}$ field. 
Moreover, the currents are strictly conserved, in the 
sense that they do not require the field equations of the undeformed Lagrangian
for the divergence to vanish. In our analysis where we imposed the upper limit 
of two derivatives in the final Lagrangian, such vertices play no role;
\item[(ii)] \underline{$a_2=0\,$, $a_1\neq 0$.} In these cases, 
the classification of vertices coming from 
all the possible $\hat{a}_1$ such that $\gamma \hat{a}_1=0$ 
has produced 
interactions of the Noether type where the currents are gauge invariant 
and conserved by virtue of the field equations of the free theory.
With our bound on the maximal number of derivatives in the Lagrangian, 
we listed many candidates $\hat{a}_1$ but only a very few of them can give 
rise to vertices. They were respectively denoted by 
$\hat{a}_1^{int}\,$, $\hat{a}_1^{_{G1}}$ and $\hat{a}_1{}^{_{G1/2}}\,$. 
The corresponding currents are, respectively, equivalent (up to trivial 
terms and on-shell vanishing expressions) to 
$J_{int.}{}^{\mu}_{\Delta} = \frac{1}{2}\,k_{a\Delta}\,F^{a}{}_{\alpha\beta}\gamma^{\alpha\beta}\gamma^\mu\,\chi\,$, 
$J^{\mu\nu}_{_{G1}} = \frac{1}{2}\,k^{ab}\,(F_a^{\mu\rho}F_{b\;\rho}^{\nu} - \tfrac{1}{4}\,\eta^{\mu\nu}\,F_a^{\alpha\beta}F_{b\alpha\beta})\,$,
and $J^{\mu\nu}_{_{G1/2}} = \partial_{\mu}\bar\chi\,\gamma^{(\mu}\partial^{\nu)}\chi\,$. 
They correspond to rigid symmetries of the free theory 
where the fields appear through gauge-invariant expressions, as 
can be read off from the expressions for 
$\hat{a}_1^{int}\,$, $\hat{a}_1^{_{G1}}$ and $\hat{a}_1{}^{_{G1/2}}\,$.
With hindsight, the corresponding interactions
result from the gauging of the corresponding rigid symmetries as we 
recall below;
\item[(iii)] \underline{$a_2\neq 0$.} 
This third case is probably the most interesting one, as it induces 
non-abelian deformations of the gauge structure of the initial 
abelian theory, 
leading to a Noether coupling where not only the current is 
non-invariant, 
but also the divergence of the current.
An example from our analysis is the current associated with the 
algebra-deforming term $a_2^{(4)}=k^{(4)}_{\Delta\Omega}\, \mathcal{C}^\star \bar{\zeta}^{[\Delta} \zeta^{\Omega]}\,$, giving rise to the 
deformation of the gauge transformations encoded in 
$a_1^{(4)} = - 4 t_{\Delta\Omega}^{(4)}\, A^{\star\mu} \bar{\psi}^{\Delta}_\mu 
\zeta^\Omega + {t}_{\Delta\Omega}\, \psi^{\star\Delta}_{\phantom{\star}\sigma} F_{\mu\nu} \gamma^{\mu\nu} \gamma^\sigma \zeta^\Omega\,$ and the conserved 
current $J^\mu_\Delta = \epsilon_{\Delta\Omega} \left( F^{\mu\nu} \Id -i \gamma^5 (*F)^{\mu\nu} \right) \psi_\nu^\Omega\,$. 
It is precisely because the corresponding $a_1$ is 
not gauge invariant that an $a_2$ term is needed, responsible 
for the deformation of the gauge structure. 
These vertices are related \cite{Julia:1980gn,Barnich:1994db,Barnich:1994mt} 
to the existence, in the initial free theory, 
of on-shell $d$-closed $n-2\,$-forms,
that are $*F^a\,$, $a=1,\ldots, n_v$ in the spin-1 case, 
equivalently conserved 2-forms $F^a\,$. 
In the spin-$\tfrac{3}{2}$ case, the components of the conserved two-forms 
are $\bar\psi^\Delta_\rho\,\gamma^{\mu\nu\rho}\,$, 
for $\Delta\in\{1,\ldots,\mathcal{N}_4\}\,$. 
In the spin-2 case, the set of forms corresponds 
to a vector-valued two-form with components 
$\Phi_{\mu\nu|}{}^{\alpha}=-\Phi_{\nu\mu|}{}^{\alpha}$ 
given in Eq. (4.11) of \cite{Boulanger:2000rq}. 
Notice the difference between spin-1 and 
spin-2: In the former case, the $n-2$ forms $* F^a$ are gauge 
invariant, whereas it is not true in the latter case since 
$\Phi_{\mu\nu|}{}^{\alpha}$ is built out of first derivatives 
of the Fierz-Pauli field. 
\end{itemize}

Actually in \cite{Julia:1980gn} it was found that an abelian local invariance 
leads to an on-shell conserved 2-form ``current'' whenever the gauge parameter 
is a closed 1-form ie when the usual gauge transformation parameter does not 
appear undifferentiated (the latter restriction is generalized by the so-called 
Killing condition below). This leads as in the global symmetry case to 
a general 
formula for the Noether-like conserved 2-form (e.g. $F^a$). 
We should take note that after a first-order abelian or non 
abelian deformation the equation of motion of the vector gauge field 
implies that the differential of the abelian $*F^a$ is proportional to 
an ordinary (on-shell closed) 3-form that corresponds to a global 
invariance by the inverse Noether theorem.
\vspace{.3cm}

Supposing we know (a) an infinitesimal deformation of 
the solution of the master equation that deforms the gauge 
transformation laws and 
(b) the existence of a Killing parameter of the initial theory, 
then one can identify a rigid symmetry of the initial theory. 
Indeed, let the initial theory $S_0[\varphi^i]=\int d^4x\, {\cal L}_0$ 
be invariant under 
$\delta_0\varphi^i = R^{(0)i}{}_\alpha\,\varepsilon^\alpha\,$. 
The existence of $W_1=\int d^4x\, ({\cal L}_1 + a_1 + a_2)$ --- where $a_2$ can 
possibly vanish --- implies that $\delta_0S_1 + \delta_1S_0=0 \,$, 
which is nothing but a rewriting of the integral of \eqref{da1ga0}.
There, one has $\delta_1 \varphi^i = R^{(1)i}{}_\alpha\,\varepsilon^\alpha\,$ 
and $S_1 = \int d^4x \,{\cal L}_1\,$. 
Killing parameters $\bar\varepsilon^\alpha$ of the initial theory 
obey  $\bar\delta_0\varphi^i \coloneqq R^{(0)i}{}_\alpha\,\bar\varepsilon^\alpha=0\,$.
Therefore, computing $\bar\delta_0S_1 + \bar\delta_1S_0$ yields 
$0 = \int d^4x\, \bar\delta_1\varphi^i\,\frac{\delta {\cal L}_0}{\delta \varphi^i}\,$, 
from which follows the rigid symmetry of the initial theory $S_0$ under 
$\bar\delta_1\varphi^i = R^{(1)i}{}_\alpha\,\bar\varepsilon^\alpha\,$. 
Therefore, the gauging of this rigid symmetry can retrospectively be viewed 
as giving the infinitesimal deformation encoded in $W_1\,$. 
\vspace{.3cm}

In the case the fields $\varphi^i$ are the vector fields $A_\mu^a\,$, 
gauging the above rigid symmetry gives 
$\delta_1 S_0 = - \int d^4x\, J^\mu_a\, \partial_{\mu}\varepsilon^a\,$
featuring the Noether current $J^\mu_a$ associated with the rigid symmetry 
$\bar\delta_1A_\mu^a = R^{(1)a}{}_{\mu b}\,\bar\varepsilon^b\,$
where $\bar\varepsilon^b\,$ are $n_v$ constants. 
On the other hand, we know that there exists $S_1$ such that 
$\delta_1 S_0 + \delta_0 S_1 = 0\,$. From the knowledge of 
$R^{(0)i}{}_\alpha\,$ for abelian vector fields, one easily 
deduces that 
$J_\mu^a + \partial^{\nu}S_{\mu\nu}^a = \frac{\delta {\cal L}_1}{\delta A^\mu_a}\,$
where $S^a_{\mu\nu} = -S^a_{\nu\mu}\,$. 
Therefore, the deformation of the action $S_0$ can be written 
$S_1 = \int d^4x\, {\cal L}_1 = \int d^4x\, \int_0^1dt\, A^\mu_a\;
\frac{\delta {\cal L}_1}{\delta A^\mu_a}(tA)$ from which 
$S_1$ appears to be proportional to the expression 
$\int d^4x\, A_\mu^a (J^\mu_a + \partial_{\nu}S^{\mu\nu}_a)\,$. 
In case ${\cal L}_1$ is cubic, one finds 
$S_1 = \frac{1}{3}\,\int d^4x\, A^\mu_a J_\mu^a\,$, showing the 
analogy with a Noether coupling, but also showing the difference in the 
overall coefficient. 
\vspace{.3cm}

In more detail, for a set of Maxwell gauge fields, the rigid 
rotational symmetry $\bar\delta_1 A_\mu^a = f^a{}_{|bc}\,A^b_\mu\,\bar\epsilon^c$ with $f_{a|bc}=-f_{b|ac}$
leads to the canonical Noether current 
$J^\mu_c = F^{\mu\nu}_a\,f^a{}_{bc}\,A_\nu^b\,$.  
On the other hand, the Yang-Mills cubic vertex reads 
${\cal L}_1 = -\frac{1}{2}\,F^{\mu\nu}\,f^a{}_{bc}\,A_\mu^b\,A_\nu^c\,$,
with totally antisymmetric constants $f_{a|bc}\,$, 
from which one derives 
$\frac{\delta {\cal L}_1}{\delta A^a_\mu} = J^\mu_a+\partial_{\nu}S^{\mu\nu}_a$
where $S_{\mu\nu}^a = - S_{\nu\mu}^a = -f^a{}_{bc}\,A_\mu^b\,A_\nu^c\,$.
The cubic action is therefore given by 
$S_1[A_\mu^a] = \frac{1}{3}\,\int d^4x\, A_\mu^a\,\frac{\delta {\cal L}_1}{\delta A^a_\mu} = \frac{1}{2}\,\int d^4x\, A_\mu^a\, J^\mu_a\,$,
where one notices the non-canonical factor $1/2\,$ due to 
the fact that the so-called ``trivial'' current 
$\partial_{\nu}S^{\mu\nu}_a$ does contribute to $S_1\,$.  
\vspace{.3cm}

The analysis proceeds similarly for the cases where $\varphi^i$
are massless spin-$\tfrac{3}{2}$ or massless spin-2 fields. 
One deduces from ${\cal L}_1$ the existence of rigid symmetries 
with on-shell conserved 
currents $J^{\mu}{}_I$ where $I=\alpha$ in the massless 
spin-2 case and $I=(A,\Delta)$ in the massless spin-$\tfrac{3}{2}$ 
case, such that 
$J^{\mu}{}_\alpha \propto \frac{\delta {\cal L}_1}{\delta h_{\mu}{}^\alpha}$ and 
$J^{\mu}{}_{A\Delta} \propto \frac{\delta {\cal L}_1}{\delta \bar\psi_{\mu}{}^{A\Delta}}\,$, 
up to the addition of ``trivial'' currents and on-shell vanishing terms.
\vspace{.3cm}

Finally, let us stress that the presence of conserved 
2 forms is tied to the presence of free abelian gauge 
fields, be them of spin 1 or 2. Moreover, in the absence of free 
abelian gauge fields, the conserved currents of Yang-Mills gauge 
models coupled to
matter fields can always be redefined so as to be gauge 
invariant as shown in \cite{Barnich:1994cq}.

\providecommand{\href}[2]{#2}\begingroup\raggedright\endgroup


\begin{thebibliography}{10}

\bibitem{Velo:1969bt}
G.~Velo and D.~Zwanziger, ``{Propagation and quantization of Rarita-Schwinger
  waves in an external electromagnetic potential},''
\href{http://dx.doi.org/10.1103/PhysRev.186.1337}{{\em Phys. Rev.} {\bf 186}
  (1969)  1337--1341}.

\bibitem{Johnson:1960vt}
K.~Johnson and E.~C.~G. Sudarshan, ``{Inconsistency of the local field theory
  of charged spin 3/2 particles},''
\href{http://dx.doi.org/10.1016/0003-4916(61)90030-6}{{\em Annals Phys.} {\bf
  13} (1961)  126--145}.

\bibitem{Adler:2015yha}
S.~L. Adler, ``{Classical Gauged Massless Rarita-Schwinger Fields},''
  \href{http://dx.doi.org/10.1103/PhysRevD.92.085022}{{\em Phys. Rev.} {\bf
  D92} (2015) no.~8, 085022},
\href{http://arxiv.org/abs/1508.03380}{{\tt arXiv:1508.03380 [hep-th]}}.

\bibitem{Adler:2015zha}
S.~L. Adler, ``{Quantized Gauged Massless Rarita-Schwinger Fields},''
  \href{http://dx.doi.org/10.1103/PhysRevD.92.085023}{{\em Phys. Rev.} {\bf
  D92} (2015) no.~8, 085023},
\href{http://arxiv.org/abs/1508.03382}{{\tt arXiv:1508.03382 [hep-th]}}.

\bibitem{Rarita:1941mf}
W.~Rarita and J.~Schwinger, ``{On a theory of particles with half integral
  spin},''
\href{http://dx.doi.org/10.1103/PhysRev.60.61}{{\em Phys. Rev.} {\bf 60} (1941)
   61}.

\bibitem{Adler:2017lki}
S.~L. Adler, M.~Henneaux, and P.~Pais, ``{Canonical Field Anticommutators in
  the Extended Gauged Rarita-Schwinger Theory},''
\href{http://arxiv.org/abs/1708.03588}{{\tt arXiv:1708.03588 [hep-th]}}.

\bibitem{Barnich:1993vg}
G.~Barnich and M.~Henneaux, ``{Consistent couplings between fields with a gauge
  freedom and deformations of the master equation},''
  \href{http://dx.doi.org/10.1016/0370-2693(93)90544-R}{{\em Phys. Lett.} {\bf
  B311} (1993)  123--129},
\href{http://arxiv.org/abs/hep-th/9304057}{{\tt arXiv:hep-th/9304057}}.

\bibitem{Batalin:1981jr}
I.~Batalin and G.~Vilkovisky, ``{Gauge Algebra and Quantization},''
  \href{http://dx.doi.org/10.1016/0370-2693(81)90205-7}{{\em Phys.Lett.} {\bf
  B102} (1981)  27--31}.

\bibitem{Batalin:1984jr}
I.~Batalin and G.~Vilkovisky, ``{Quantization of Gauge Theories with Linearly
  Dependent Generators},'' \href{http://dx.doi.org/10.1103/PhysRevD.28.2567,
  10.1103/PhysRevD.30.508}{{\em Phys.Rev.} {\bf D28} (1983)  2567--2582}.

\bibitem{Berends:1984rq}
F.~A. Berends, G.~J.~H. Burgers, and H.~van Dam, ``{On the theoretical problems
  in constructing interactions involving higher spin massless particles },''
\href{http://dx.doi.org/10.1016/0550-3213(85)90074-4}{{\em Nucl. Phys.} {\bf
  B260} (1985)  295}.

\bibitem{Boulanger:2000rq}
N.~Boulanger, T.~Damour, L.~Gualtieri, and M.~Henneaux, ``{Inconsistency of
  interacting, multigraviton theories},''
  \href{http://dx.doi.org/10.1016/S0550-3213(00)00718-5}{{\em Nucl. Phys.} {\bf
  B597} (2001)  127--171},
\href{http://arxiv.org/abs/hep-th/0007220}{{\tt arXiv:hep-th/0007220}}.

\bibitem{Henneaux:2012wg}
M.~Henneaux, G.~Lucena~Gomez, and R.~Rahman, ``{Higher-Spin Fermionic Gauge
  Fields and Their Electromagnetic Coupling},''
  \href{http://dx.doi.org/10.1007/JHEP08(2012)093}{{\em JHEP} {\bf 1208} (2012)
   093},
\href{http://arxiv.org/abs/1206.1048}{{\tt arXiv:1206.1048 [hep-th]}}.

\bibitem{Barnich:2017nty}
G.~Barnich, N.~Boulanger, M.~Henneaux, B.~Julia, V.~Lekeu, and A.~Ranjbar,
  ``{Deformations of vector-scalar models},''
\href{http://arxiv.org/abs/1712.08126}{{\tt arXiv:1712.08126 [hep-th]}}.

\bibitem{Bizdadea:2015yip}
C.~Bizdadea and S.-O. Saliu, ``{Gauge-invariant massive BF models},''
  \href{http://dx.doi.org/10.1140/epjc/s10052-016-3913-3}{{\em Eur. Phys. J.}
  {\bf C76} (2016) no.~2, 65},
\href{http://arxiv.org/abs/1511.04684}{{\tt arXiv:1511.04684 [hep-th]}}.

\bibitem{Henneaux:2013gba}
M.~Henneaux, G.~Lucena~G{\'o}mez, and R.~Rahman, ``{Gravitational Interactions
  of Higher-Spin Fermions},''
  \href{http://dx.doi.org/10.1007/JHEP01(2014)087}{{\em JHEP} {\bf 01} (2014)
  087},
\href{http://arxiv.org/abs/1310.5152}{{\tt arXiv:1310.5152 [hep-th]}}.

\bibitem{Ferrara:1976um}
S.~Ferrara, J.~Scherk, and P.~van Nieuwenhuizen, ``{Locally Supersymmetric
  Maxwell-Einstein Theory},''
\href{http://dx.doi.org/10.1103/PhysRevLett.37.1035}{{\em Phys. Rev. Lett.}
  {\bf 37} (1976)  1035}.

\bibitem{Freedman:1976aw}
D.~Z. Freedman and A.~K. Das, ``{Gauge Internal Symmetry in Extended
  Supergravity},''
\href{http://dx.doi.org/10.1016/0550-3213(77)90041-4}{{\em Nucl.Phys.} {\bf
  B120} (1977)  221}.

\bibitem{Fradkin:1976xz}
E.~S. Fradkin and M.~A. Vasiliev, ``{SO(2) supergravity with minimal
  electromagnetic interaction, 1.5-order formalism and the explicit form of the
  structure coefficients},''
{\em Lebedev Institute Preprint 197 (1976)} (1976)  .

\bibitem{Ferrara:1976fu}
S.~Ferrara and P.~van Nieuwenhuizen, ``{Consistent Supergravity with Complex
  Spin 3/2 Gauge Fields},''
\href{http://dx.doi.org/10.1103/PhysRevLett.37.1669}{{\em Phys. Rev. Lett.}
  {\bf 37} (1976)  1669}.

\bibitem{Freedman:1976nf}
D.~Z. Freedman, ``{SO(3) Invariant Extended Supergravity},''
\href{http://dx.doi.org/10.1103/PhysRevLett.38.105}{{\em Phys. Rev. Lett.} {\bf
  38} (1977)  105}.

\bibitem{Boulanger:2001wq}
N.~Boulanger and M.~Esole, ``{A Note on the uniqueness of D = 4, N=1
  supergravity},'' \href{http://dx.doi.org/10.1088/0264-9381/19/8/304}{{\em
  Class. Quant. Grav.} {\bf 19} (2002)  2107--2124},
\href{http://arxiv.org/abs/gr-qc/0110072}{{\tt arXiv:gr-qc/0110072 [gr-qc]}}.

\bibitem{Townsend:1977qa}
P.~K. Townsend, ``{Cosmological Constant in Supergravity},''
\href{http://dx.doi.org/10.1103/PhysRevD.15.2802}{{\em Phys. Rev.} {\bf D15}
  (1977)  2802--2804}.

\bibitem{Barnich:1994db}
G.~Barnich, F.~Brandt, and M.~Henneaux, ``{Local BRST cohomology in the
  antifield formalism. 1. General theorems},''
  \href{http://dx.doi.org/10.1007/BF02099464}{{\em Commun. Math. Phys.} {\bf
  174} (1995)  57--92},
\href{http://arxiv.org/abs/hep-th/9405109}{{\tt arXiv:hep-th/9405109}}.

\bibitem{Barnich:1994mt}
G.~Barnich, F.~Brandt, and M.~Henneaux, ``{Local BRST cohomology in the
  antifield formalism. II. Application to Yang-Mills theory},''
  \href{http://dx.doi.org/10.1007/BF02099465}{{\em Commun. Math. Phys.} {\bf
  174} (1995)  93--116},
\href{http://arxiv.org/abs/hep-th/9405194}{{\tt arXiv:hep-th/9405194}}.

\bibitem{Henneaux:1992ig}
M.~Henneaux and C.~Teitelboim, ``{Quantization of gauge systems},''. Princeton,
  USA: Univ. Pr. (1992) 520 p.

\bibitem{Gomis:1994he}
J.~Gomis, J.~Paris, and S.~Samuel, ``{Antibracket, antifields and gauge theory
  quantization},'' \href{http://dx.doi.org/10.1016/0370-1573(94)00112-G}{{\em
  Phys.Rept.} {\bf 259} (1995)  1--145},
\href{http://arxiv.org/abs/hep-th/9412228}{{\tt arXiv:hep-th/9412228
  [hep-th]}}.

\bibitem{LucasMasterThesis}
L.~Traina, ``Th\'eorie de supergravit\'e \'etendue $n=2$ en dimension 4,''
  Master's thesis, Universit\'e de Mons -- UMONS,
  http://hosting.umons.ac.be/php/mecagrav/english/traina.php, 2016.

\bibitem{Pilch:1984aw}
K.~Pilch, P.~van Nieuwenhuizen, and M.~F. Sohnius, ``{De Sitter Superalgebras
  and Supergravity},''
\href{http://dx.doi.org/10.1007/BF01211046}{{\em Commun. Math. Phys.} {\bf 98}
  (1985)  105}.

\bibitem{Bergshoeff:2015tra}
E.~A. Bergshoeff, D.~Z. Freedman, R.~Kallosh, and A.~Van~Proeyen, ``{Pure de
  Sitter Supergravity},'' \href{http://dx.doi.org/10.1103/PhysRevD.93.069901,
  10.1103/PhysRevD.92.085040}{{\em Phys. Rev.} {\bf D92} (2015) no.~8, 085040},
  \href{http://arxiv.org/abs/1507.08264}{{\tt arXiv:1507.08264 [hep-th]}}.
[Erratum: Phys. Rev.D93,no.6,069901(2016)].

\bibitem{Barnich:1993pa}
G.~Barnich, M.~Henneaux, and R.~Tatar, ``{Consistent interactions between gauge
  fields and the local BRST cohomology: The Example of Yang-Mills models},''
  \href{http://dx.doi.org/10.1142/S0218271894000149}{{\em Int.J.Mod.Phys.} {\bf
  D3} (1994)  139--144},
\href{http://arxiv.org/abs/hep-th/9307155}{{\tt arXiv:hep-th/9307155
  [hep-th]}}.

\bibitem{Cremmer:1978hn}
E.~Cremmer, B.~Julia, J.~Scherk, S.~Ferrara, L.~Girardello, and P.~van
  Nieuwenhuizen, ``{Spontaneous Symmetry Breaking and Higgs Effect in
  Supergravity Without Cosmological Constant},''
\href{http://dx.doi.org/10.1016/0550-3213(79)90417-6}{{\em Nucl. Phys.} {\bf
  B147} (1979)  105}.

\bibitem{Cremmer:1982en}
E.~Cremmer, S.~Ferrara, L.~Girardello, and A.~Van~Proeyen, ``{Yang-Mills
  Theories with Local Supersymmetry: Lagrangian, Transformation Laws and
  SuperHiggs Effect},''
\href{http://dx.doi.org/10.1016/0550-3213(83)90679-X}{{\em Nucl. Phys.} {\bf
  B212} (1983)  413}.

\bibitem{Cremmer:1984hj}
E.~Cremmer, C.~Kounnas, A.~Van~Proeyen, J.~P. Derendinger, S.~Ferrara,
  B.~de~Wit, and L.~Girardello, ``{Vector Multiplets Coupled to N=2
  Supergravity: SuperHiggs Effect, Flat Potentials and Geometric Structure},''
\href{http://dx.doi.org/10.1016/0550-3213(85)90488-2}{{\em Nucl. Phys.} {\bf
  B250} (1985)  385--426}.

\bibitem{Freedman:1976xh}
D.~Z. Freedman, P.~van Nieuwenhuizen, and S.~Ferrara, ``{Progress Toward a
  Theory of Supergravity},''
\href{http://dx.doi.org/10.1103/PhysRevD.13.3214}{{\em Phys. Rev.} {\bf D13}
  (1976)  3214--3218}.

\bibitem{Deser:1976eh}
S.~Deser and B.~Zumino, ``{Consistent Supergravity},''
\href{http://dx.doi.org/10.1016/0370-2693(76)90089-7}{{\em Phys. Lett.} {\bf
  62B} (1976)  335}.

\bibitem{VanProeyen:1999ni}
A.~Van~Proeyen, ``{Tools for supersymmetry},'' {\em Ann. U. Craiova Phys.} {\bf
  9} (1999) no.~I, 1--48,
\href{http://arxiv.org/abs/hep-th/9910030}{{\tt arXiv:hep-th/9910030
  [hep-th]}}.

\bibitem{Henneaux:1997bm}
M.~Henneaux, ``{Consistent interactions between gauge fields: The cohomological
  approach},'' {\em Contemp. Math.} {\bf 219} (1998)  93,
\href{http://arxiv.org/abs/hep-th/9712226}{{\tt arXiv:hep-th/9712226}}.

\bibitem{Julia:1980gn}
B.~Julia, ``{Effective Gauge Fields and Generalized Noether Theorem},'' in {\em
  {4th Workshop on Current Problems in High-Energy Particle Theory Bad Honnef,
  Germany, June 2-4, 1980}}, pp.~295--313.
\newblock
1980.
\newblock

\bibitem{Barnich:1994cq}
G.~Barnich, F.~Brandt, and M.~Henneaux, ``{Conserved currents and gauge
  invariance in Yang-Mills theory},''
  \href{http://dx.doi.org/10.1016/0370-2693(95)00011-9}{{\em Phys. Lett.} {\bf
  B346} (1995)  81--86},
\href{http://arxiv.org/abs/hep-th/9411202}{{\tt arXiv:hep-th/9411202
  [hep-th]}}.

\end{thebibliography}

\end{document}